\RequirePackage{fixltx2e}
\documentclass[aip,jcp,reprint,floatfix]{revtex4-1}

\usepackage{amsmath,amsfonts} %
\usepackage[acronym]{glossaries} %
\usepackage{graphicx} %
\usepackage{color} %
\usepackage[breaklinks, %
colorlinks, %
linkcolor=blue, %
citecolor=blue, %
urlcolor=blue]{hyperref} %
\usepackage{braket} %
\newcommand{\braOket}[3]{\ensuremath{\left\langle #1\left|#2\right|#3\right\rangle}}
\newcommand{\norm}[1]{||#1||} %
\newcommand{\eq}[1]{Eq.~(\ref{#1})} %
\newcommand{\bea}{\begin{eqnarray}}
\newcommand{\eea}{\end{eqnarray}}
\newcommand{\mat}[1]{\ensuremath{\boldsymbol{#1}}}
\newcommand{\op}[1]{\ensuremath{\hat{#1}}}

\newcommand{\D}{\mathcal{D}}
\newcommand{\ie}{\emph{i.e.} }
\ifpdf %
 \fi

\newacronym{NAC}{NAC}{non-adiabatic coupling} %
\newacronym{DC}{DC}{derivative coupling} %
\newacronym{GD}{GD}{gradient difference} %
\newacronym{CI}{CI}{conical intersection} %
\newacronym{MECI}{MECI}{minimum energy conical intersection} %
\newacronym{BS}{BS}{branching space} %
\newacronym{IS}{IS}{intersection space}
\newacronym{CG}{CG}{composed gradient}
\newacronym{UBS}{UBS}{update of the branching space}
\newacronym{LM}{LM}{Lagrange multipliers}
\newacronym{ALM}{ALM}{approximate \gls{LM}}
\newacronym{SLM}{SLM}{single \gls{LM}}
\newacronym{BFGS}{BFGS}{Broyden-Fletcher-Goldfarb-Shanno}
\newacronym{RMS}{RMS}{root mean square}
\newacronym{RMSD}{RMSD}{root mean square deviation}
\newacronym{LVC}{LVC}{linear vibronic coupling} %
\newacronym{DOF}{DOF}{degrees of freedom} %
\newacronym{PES}{PES}{potential energy surface} %
\newacronym{FC}{FC}{Franck-Condon} %
\newacronym{CASSCF}{CASSCF}{complete active space self consistent field}
\newacronym{TDDFT}{TDDFT}{time dependent density functional theory}
\newacronym{MSCASPT2}{MS-CASPT2}{multi-state complete active space second order perturbation theory}

\begin{document}

\author{Juan Sanz García} %
\affiliation{Univ Gustave Eiffel, Univ Paris Est Creteil, CNRS, UMR 8208, MSME, F-77454 Marne-la-Vall\'ee, France} %

\author{Rosa Maskri} %
\affiliation{Univ Gustave Eiffel, Univ Paris Est Creteil, CNRS, UMR 8208, MSME, F-77454 Marne-la-Vall\'ee, France} %

\author{Alexander Mitrushchenkov} %
\affiliation{Univ Gustave Eiffel, Univ Paris Est Creteil, CNRS, UMR 8208, MSME, F-77454 Marne-la-Vall\'ee, France} %

\author{Lo{\"i}c Joubert-Doriol}
\email{loic.joubert-doriol@univ-eiffel.fr}
\affiliation{Univ Gustave Eiffel, Univ Paris Est Creteil, CNRS, UMR 8208, MSME, F-77454 Marne-la-Vall\'ee, France} %

\title{Optimizing Conical Intersections Without Explicit Use of Non-Adiabatic Couplings}

\begin{abstract}
We present two alternative methods for optimizing \gls{MECI} molecular geometries without knowledge of the \gls{DC}.
These methods are based on the utilization of \acrlong{LM}: i) one method uses an approximate calculation of the \gls{DC}, while the other ii) do not require the \gls{DC}.
Both methods use the fact that information of the \gls{DC} is contained in the Hessian of the squared energy difference.
Tests done on a set of small molecular systems, in comparison with other methods, show the ability of the proposed methods to optimize \glspl{MECI}.
Finally, we apply the methods to the furimamide molecule, to optimize and characterize its $S_1$/$S_2$ \gls{MECI}, and to optimizing the $S_0$/$S_1$ \gls{MECI} of the silver trimer.
\end{abstract}

%\date{\today}

\maketitle

\glsresetall

%%%%%%%%%%%%%%%%%%%%%%%%%%%%%%%%%%%%%%%%%%%%%%
\section{Introduction}
\label{sec:introduction}

% Introduction on CIs' importance
\Glspl{CI} are well-known features of molecular adiabatic \glspl{PES} responsible for non-adiabatic dynamics in photophysics and photochemistry.~\cite{Yarkony:2001/jpca/6277,Migani:2004/271,Persico:2018,Matsika:2021/cr/9407}
These features, i.e. \glspl{CI}, are nuclear geometries for which adiabatic electronic states are degenerate. 
In these geometries, the \acrlongpl{NAC} become singular,~\cite{Migani:2004/271,Baer:2006,Persico:2018} which allows non-radiative population transfer between adiabatic electronic states.
Accounting for possible non-radiative transitions in molecular simulations is the topic of intense research.~\cite{Richings:2015/irpc/269,Wang:2016/jpcl/2100,Freixas:2021/jpcl/2970,Maskri:2022/ptrsa/20200379,Lassmann:2023/tca/1}
Another interesting aspect is the appearance of a topological phase in the adiabatic electronic states,~\cite{Mead:1979/jcp/2284,Berry:1987/rspa/31} which play an important role even in low energy dynamics in the vicinity of \glspl{CI}.~\cite{Ryabinkin:2017/acr/1785,Joubert:2017/cc/7365}
Hence, prior location of \glspl{CI} geometries is essential for a better understanding of non-adiabatic dynamics in both photochemical and photophysical processes.

% Basic info to optimize MECIs
The set of \gls{CI} geometries forms a ($\D-2$)-dimensional space, the \gls{IS}, where $\D$ is the dimensionality of the internal nuclear coordinates space ($\D=3N-6$ for $N$ nuclei).~\cite{Yarkony:2004/41}
Among all geometries of the \gls{IS}, the \gls{MECI} is often sought because the region of the \gls{PES} containing this stationary point is the most accessible for the system.
Finding a \gls{MECI} point between two electronic adiabatic \glspl{PES} is equivalent to minimizing the energy of both states while constraining their energy difference to be equal to zero.
It can be shown that two directions of collective nuclear motion lift the degeneracy at a \gls{CI} geometry at first order in the nuclear displacements.~\cite{Yarkony:2004/41}
These directions are given by the \gls{GD} of the adiabatic energies and the \gls{DC} vector, and together, they
define the $2$-dimensional \gls{BS} which is complementary to the \gls{IS}.
Therefore, in their most straightforward implementations, \gls{MECI} optimization algorithms require the knowledge of the \gls{GD} and \gls{DC} vectors in order to constrain the involved electronic states to remain degenerate.~\cite{Manaa:1993/jcp/5251,Bearpark:1994/cpl/269}
Moreover, it is possible to explore the \gls{IS} by employing a similar strategy.~\cite{Sicilia:2008/jctc/257}

% Alternatives when DC cannot be calculated
Computing the \gls{GD}, analytically or numerically, is possible as long as adiabatic energies can be computed.
This is not the case for a direct evaluation of the \gls{DC}, which requires the knowledge of the wavefunction, for example using the ``off-diagonal'' Hellmann-Feynman-like theorem~\cite{Persico:2018} and analytic or numerical differentiation.
Hence, a problem arises when the wavefunctions are not available.
In wavefunction-based electronic structure methods, the analytic evaluation of the \gls{DC} can become computationally expensive or difficult to implement and, thus, it may not be available for numerically complex electronic structure methods.
In such cases, one must employ alternative \gls{MECI} optimization methods in which the \gls{DC} is not required.
Penalty functions have been defined to overcome the absence of the \gls{DC} vector.~\cite{Ciminelli:2004/cej/2327,Levine:2008/jpcb/405}
However, algorithms using these penalty functions suffer from slow convergence and the exact degeneracy can never be reached.~\cite{Keal:2007/tca/837}
Yet another very powerful method, developed by Morokuma and coworkers,~\cite{Maeda:2010/jctc/1538} enables \gls{MECI} optimization without the need for evaluating the \gls{DC} vector.
We shall name this method the \gls{UBS}.
It is worth mentioning that, in contrast to the approach using penalty functions, the Morokuma's method converges to an exact \gls{MECI}.
This method is based on the \gls{CG} approach,~\cite{Bearpark:1994/cpl/269} which requires the construction of the projector on the \gls{BS}.
Constructing the \gls{BS} projector can be achieved using the \gls{DC} vector.
In such a case, the \gls{DC} must be computed in a first place.
However, computing the \gls{DC} vector is not necessary if the projector can be directly evaluated.
This is exactly what the \gls{UBS} method does: the projector is approximately calculated along the optimization by using an updating procedure, while the \gls{DC} vector is never computed.

% Difficulties in CG and UBS
The \gls{UBS} method relies on the fact that variations with respect to nuclear \gls{DOF} of the adiabatic energy difference, $\Omega=E_l-E_k$, between the two intersecting adiabatic states, $\varphi_k$ and $\varphi_l$, contain information about the \gls{DC} vector.~\cite{Yarkony:2001/jpca/6277}
Hence, the \gls{BS} can be defined, without the need of the \gls{DC}, using the variations of $\Omega$ with respect to nuclear \gls{DOF}.
Indeed, one can obtain an approximation of the energy difference in the vicinity of the \gls{CI} by applying a first order perturbative analysis in the nuclear displacements:~\cite{Yarkony:2001/jpca/6277,Yarkony:2004/41}
\bea\label{eq:DE@CI}
\Omega & = & \sqrt{[\mat d^t(\mat X-\mat X_{CI})]^2+4[\mat g^t(\mat X-\mat X_{CI})]^2},
\eea
where $\Omega=E_l-E_k$ is the energy difference between the two intersecting adiabatic states, $\mat d=\mat\nabla (E_l-E_k)$ is the \gls{GD}, $\mat g=\bra{\varphi_k}{\mat\nabla\op H_e}\ket{\varphi_l}$ is the \gls{DC}, $\op H_e$ is the electronic Hamiltonian (\ie $\op H_e\ket{\varphi_k}=E_k\ket{\varphi_k}$), and $\mat X$ is a vector of rectilinear nuclear coordinates, $\mat X_{CI}$ being the position of the \gls{CI} in this coordinates' system.
In \eq{eq:DE@CI} and in the rest of this manuscript, boldface symbols represent matrices and vectors, the superscript ``$t$'' indicates the transpose of a matrix (or a vector), the nabla symbol ``$\mat\nabla$'' means the gradient with respect to nuclear coordinates, and the intrinsic dependence of all the quantities on the nuclear coordinates will be assumed when not written explicitly.
In the \gls{UBS} method, the information of the \gls{DC} vector that is contained in the energy difference is directly extracted from the \gls{GD} obtained at each geometry along the \gls{CI} optimization.
The \gls{UBS} method relies on two key points, i) the initial guess of the \gls{BS} projector and ii) the successful application of the approximate update scheme. The approximate construction of the projector can introduce difficulties in the global convergence properties of the CI minimization.
This is clearly a disadvantage compared to \gls{CG} methods where no approximate quantities are used.
Another difficulty is rooted in the \gls{CG} method itself, which aims at solving a vector equation, where the vector does not derive from a scalar function.
Hence, applying the quasi-Newton algorithm to solve this vector equation leads to numerical problems.~\cite{Toniolo:2002/jpca/4679,Ruiz-Barragan/2013/jctc/1433}
There exists a composed-step algorithm that aims at solving this problem,~\cite{Ruiz-Barragan/2013/jctc/1433} and could in principle be combined with the \gls{UBS} method.

% alternative to the CG method: LM
Using \gls{LM} can be a good alternative to the \gls{CG} method.
The \gls{LM} technique imposes exact constraints on the energy difference while minimizing the energy.~\cite{Manaa:1993/jcp/5251} 
This alternative method generally converges faster to a \gls{CI}.~\cite{Keal:2007/tca/837}
Combined with the \gls{LM} technique, the quasi-Newton algorithm can be used to approximate the second derivative quantities.
In order to apply the quasi-Newton algorithm, one needs to avoid the undefined \gls{GD} at the \gls{CI} (due to the square root).
This can be achieved by applying constraints on the two following functions that appear inside the square root in \eq{eq:DE@CI}:
\bea
\Delta \equiv (\mat X-\mat X_{CI})^t\mat d & = & 0, \label{eq:defDelta}\\
\Gamma \equiv 2(\mat X-\mat X_{CI})^t\mat g & = & 0. \label{eq:defGamma}
\eea
Note that these functions resemble the energy difference and the coupling in a \acrlong{LVC} model.~\cite{Yarkony:2004/41,Koppel:2004/175}
Applying \gls{LM} requires the derivative of $\Gamma$, and thus, the calculation of the \gls{DC}.
Despite this apparent difficulty, we have used the \gls{LM} technique to devise two new methods for the minimization of \glspl{CI} without computing the \gls{DC} vector.
Our developments are strongly related to the pioneering work of K\"oppel and Schubert,~\cite{Koppel:2006/mp/1069}. 
They developed a method using only the energy to obtain quantities related to the \gls{DC} by computing the squared energy difference Hessian.
This approach is also strongly related to the energy-based diabatization method, where no information of the wavefunction is needed.~\cite{Koppel:2004/175}
After this work, a similar methodology has been applied to estimate the \gls{DC},~\cite{Kammeraad:2016/jpcl/5074,Gonon:2017/jcp/114114,Baeck:2017/jcp/064107,Chen:2023/mol/4222,Vandaele:2024/jctc/} and is, in fact, also employed in the \gls{UBS} method.~\cite{Maeda:2010/jctc/1538}
Indeed, at a \gls{CI} between states $k$ and $l$, it can be demonstrated that (see App.~\ref{app:ddO2})
\bea\label{eq:ddO2}
\mat g \mat g^t & = & \frac{1}{8}\mat\nabla\mat\nabla^t\Omega^2 - \frac{1}{4} \mat d \mat d^t.
\eea
It must be emphasized that \eq{eq:ddO2} is exact at \glspl{CI} but approximate if $E_k\ne E_l$.
However, at \gls{CI} optimization convergence, electronic energies are numerically degenerate, and thus, \eq{eq:ddO2} applies.

\begin{figure}
  \centering
  \begin{tabular}{cc}
  \begin{tabular}{cc}
    \vspace{-0.2cm}
    \includegraphics[width=0.1\textwidth]{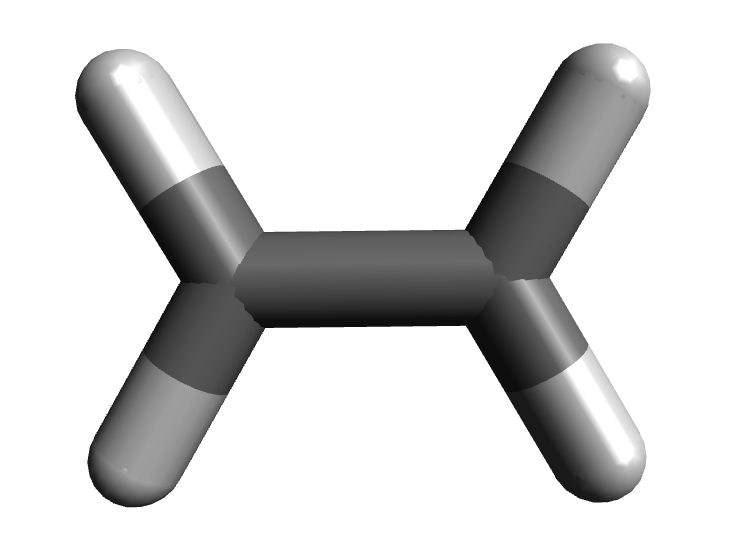}&\includegraphics[width=0.15\textwidth]{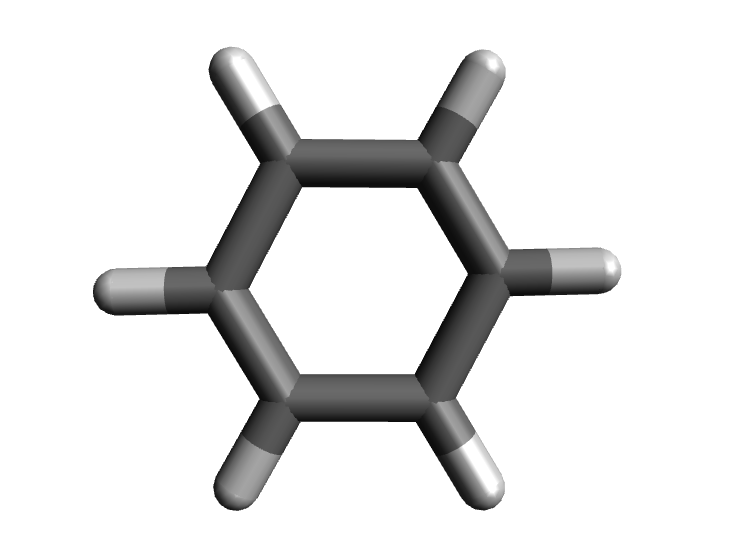}\\
    Ethylene & Benzene \\
    \vspace{-0.2cm}
    \includegraphics[width=0.1\textwidth]{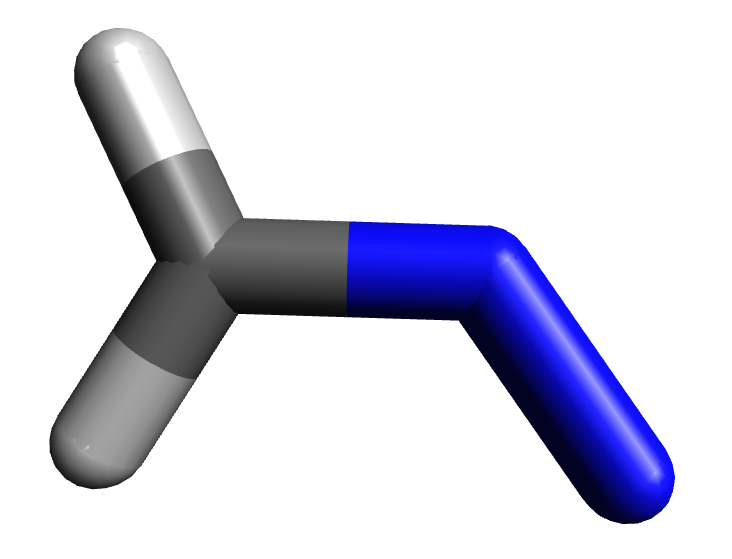}&\includegraphics[width=0.1\textwidth]{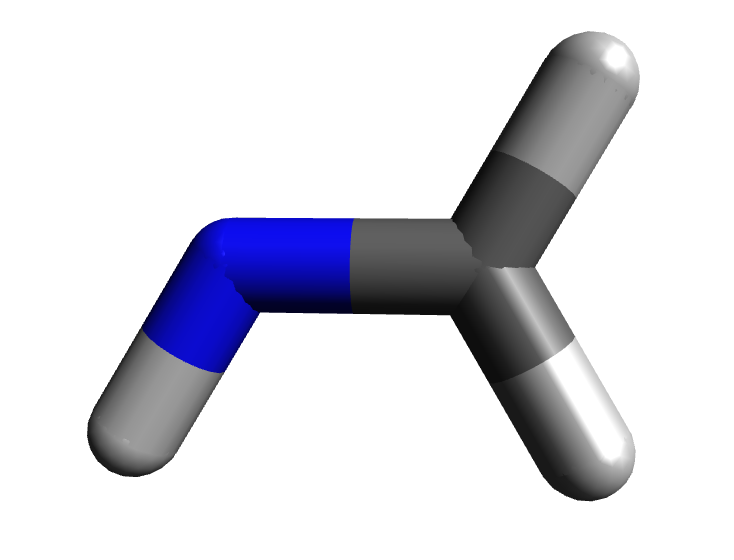}\\
    Diazomethane & Methaniminium \\
    \includegraphics[width=0.3\textwidth]{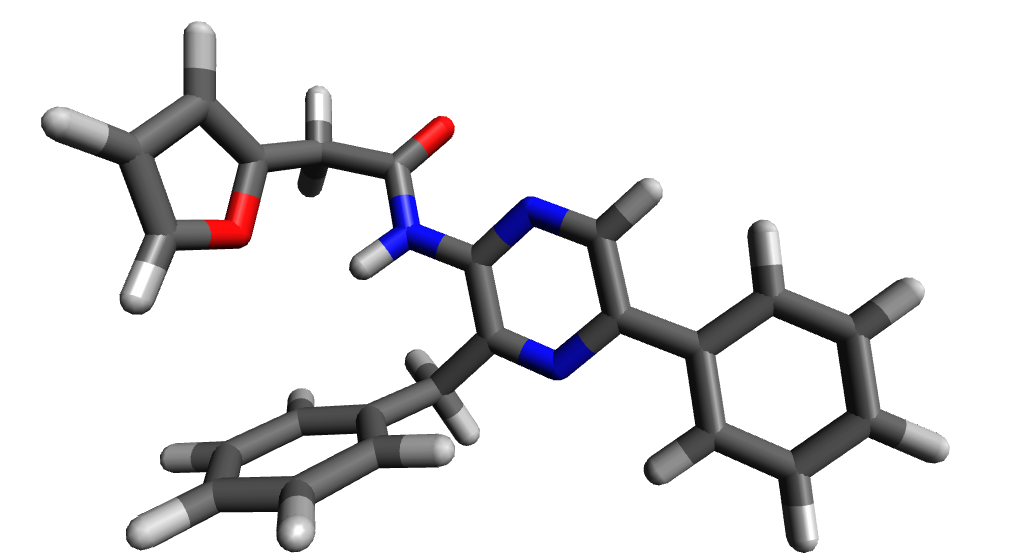}&\hspace{-0.4cm}\includegraphics[width=0.1 \textwidth]{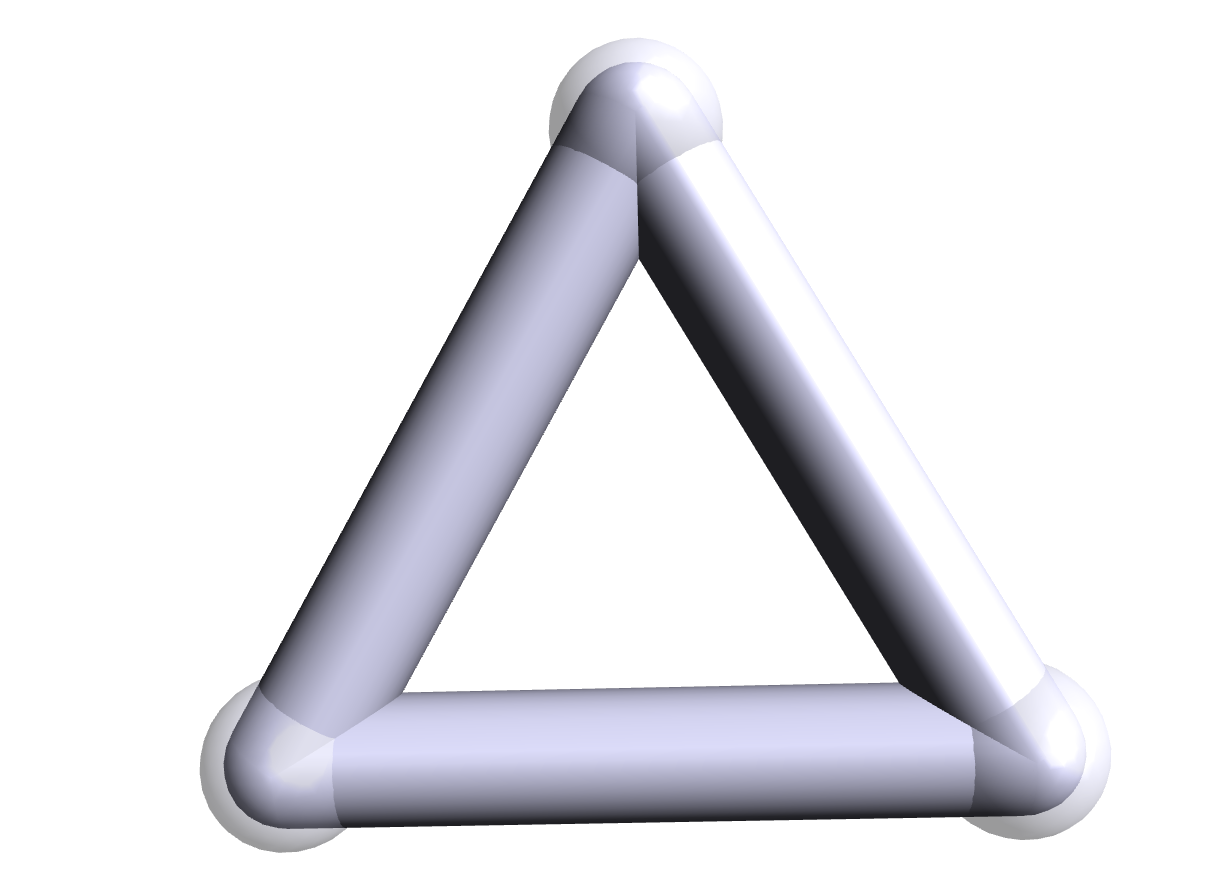}\\
    Furimamide & Ag$_3$ \\[0.25cm]
  \end{tabular}\\
  \end{tabular}
  \caption{Molecules for which \acrlongpl{MECI} are optimized in this study.}
  \label{fig:molecules}
\end{figure}

% New methods of this paper
Using the relation given by \eq{eq:ddO2}, we have devised two methods: 
i) an approximate \gls{DC} is computed and used in combination with the \gls{LM} approach: this method will be dubbed \gls{ALM}, and 
ii) without using the functions $\Delta$ and $\Gamma$ a constraint is imposed on the squared energy difference, $\Omega^2=0$, which does not show a singular behavior at the \gls{CI}: since only one constraint is employed, this method will be dubbed \gls{SLM}.
The \gls{ALM} method is in fact very similar in spirit to the recently developed method described in Ref.~\onlinecite{Fdez.Galvan:2023/jctc/3418}, 
Such diabatic models have been commonly built along ``trajectories'', see for example Refs.~\cite{Zhu:2009/jcp/234108,Zhu:2010/jcp/104101}.
the main difference being that here we employ a local diabatic model to estimate the \gls{DC}, that we do  not calculate, rather than optimizing the \gls{MECI} in the diabatic representation with the help of the calculated \gls{DC}.
We have tested these methods on simple molecular systems at low computational cost, where the \gls{DC} can easily be calculated, and the \gls{LM} method can be used as a reference result.
As a comparison, we have also run the same optimizations using the \gls{CG} and the \gls{UBS} methods.
We have chosen the following molecular systems (see Fig.~\ref{fig:molecules}): ethylene,~\cite{Barbatti:2004/jcp/2004} benzene,~\cite{Palmer:1993/jacs/673} methaniminium,~\cite{Barbatti:2006/mp/1053} and diazomethane,~\cite{Yamamoto:1994/jacs/2064,Page:2003/jcc/298} which all exhibit well-known \gls{CI}, and were used before to test \gls{MECI} algorithms (see Ref.~\onlinecite{Keal:2007/tca/837}).
In order to show the real applicability of the developed methods, we also have applied both of the methods, \gls{ALM} and \gls{SLM}, to two molecular systems that were recently the subject of published research works and for which the non-adiabatic coupling calculations imply a significant computation time.
These systems are:
1) The furimamide molecule (C$_{24}$N$_3$O$_2$H$_{19}$), which is a remarkable nanoluciferase's ligand involved in very efficient bioluminescence processes,~\cite{Hall:2012/acscb/1848} for which we optimize a recently discovered S$_1$/S$_2$ \gls{CI}~\cite{Sahihi:2020/jpcb/2539} at the \gls{TDDFT} level since it is a rather large molecular system.
2) The silver trimer (Ag$_{3}$), which is a model for the conical intersection optimization of silver and copper cluster study,~\cite{Mitrushchenkov:2023/pcp/e202300317} for which we optimize the S$_0$/S$_1$ \gls{CI} at the \gls{MSCASPT2} level of theory.

% Plan
The rest of the paper is organized as follows.
The methods \gls{ALM} and \gls{SLM} are described in Sec.~\ref{sec:theory}.
Computational details are given in Sec.~\ref{sec:details}.
Section~\ref{sec:test} is devoted to testing these methods on simple molecular systems as a proof of concept.
In section~\ref{sec:application} we apply the new methods to the optimization of the \gls{CI} of the furimamide molecule.
The last section, Sec.~\ref{sec:conclusion} concludes this paper.

%%%%%%%%%%%%%%%%%%%%%%%%%%%%%%%%%%%%%%%%%%%%%%%%%%%%%%%%%%%%%%%%%%%%%%
\section{Theoretical developments}
\label{sec:theory}

In this section, we describe the details of the proposed \gls{MECI} optimization methods \gls{ALM} and \gls{SLM}.
We assume that the electronic problem can be solved using quantum chemistry methods, so that the set of electronic adiabatic states $\{\ket{\varphi_k}\}$ and associated adiabatic energies $E_k$ are known.
We further assume that energy gradients can also be obtained.

Before describing our methods, we first need to define some key quantities that will be used in the definition of a \gls{MECI}.
A \gls{MECI} is characterized by a minimal energy within the $(\mathcal{D}-2)$-dimensional subspace where $E_k=E_l$, the \glsfirst{IS}.
The \gls{BS} is defined as the complementary $2$-dimensional space.
Hence, if we can define projectors on each of these subspaces by $\mat P_{IS}$ and $\mat P_{BS}$, we must have $\mat P_{IS}+\mat P_{BS}=\mat 1_\mathcal{D}$ where $\mat 1_\mathcal{D}$ is the $\mathcal{D}$-dimensional unit matrix.
The branching space can also be defined as the $2$-dimensional subspace which contains the directions given by \gls{GD} and \gls{DC}.
Thus, we can express the \gls{BS} projector using the following defintions
\bea
\mat P_{BS} & = & \mat B(\mat B^t\mat B)^{-1}\mat B^t, \label{eq:Pbs}\\
\mat B & = & \begin{pmatrix}\mat d&\mat g\end{pmatrix}. \label{eq:B}
\eea
The conditions for reaching a \gls{MECI} can then be rewritten as $\Omega=0$ and $\norm{\mat P_{IS}\mat\nabla\Sigma}=0$, where $\Sigma=E_k+E_l$ is the sum of the energies.
Note that we have the relation $\mat\nabla\Sigma=2\mat\nabla E_k+\mat d=2\mat\nabla E_l-\mat d$, which becomes $\mat P_{IS}\mat\nabla\Sigma=2\mat P_{IS}\mat\nabla E_k=2\mat P_{IS}\mat\nabla E_l$ because by construction $\mat P_{BS}\mat d=\mat d$.
Hence, the three conditions $\norm{\mat P_{IS}\mat\nabla\Sigma}=0$, $\norm{\mat P_{IS}\mat\nabla E_k}=0$, and $\norm{\mat P_{IS}\mat\nabla E_l}=0$ are equivalent.

\subsection{The approximate method: \protect\gls{ALM}}
\label{sub:ALM}

The \gls{ALM} method is based on the \gls{LM} approach, which minimizes the energy while constraining the energy difference to zero.
Due to the root square appearing in \eq{eq:DE@CI}, the gradient of $\Omega=E_k-E_l$ is not defined at the \gls{CI}, and we cannot use for example a Taylor expansion to approximate $\Omega$ in the vicinity of the \gls{CI}.
Instead, we Taylor expand $\Omega^2$, whose gradient is well-defined at the \gls{CI}, about the nuclear geometry $\mat X_n$:
\bea\label{eq:Taylor}
\Omega^2 & \approx & (\Omega^{(n)})^2 \equiv \Omega_n^2 + (\mat X - \mat X_n)^t\mat\nabla\Omega^2\Big|_{\mat X_n} 
\nonumber\\&&
+ \frac{1}{2}(\mat X - \mat X_n)^t \mat\nabla\mat\nabla^t\Omega^2\Big|_{\mat X_n}(\mat X - \mat X_n), 
\eea
where the ``$\equiv$'' symbol defines $\Omega^{(n)}$.
In \eq{eq:Taylor} and the rest of the paper, the index $n$ will be used to denote the nuclear geometry $\mat X_n$ at which the corresponding quantities are calculated.
Isolating and substituting the Hessian $\mat\nabla\mat\nabla^t\Omega^2$ from \eq{eq:ddO2} into \eq{eq:Taylor}, and expanding $\mat\nabla\Omega^2=2\Omega\mat\nabla\Omega$, we obtain, after factorization
\bea\label{eq:appDE}
(\Omega^{(n)})^2 & = & (\Delta^{(n)})^2 + (\Gamma^{(n)})^2.
\eea
where we define the following functions about the geometry $\mat X_n$
\bea
\Delta^{(n)} & \equiv & \Omega_n + (\mat X - \mat X_n)^t\mat d_n, \\
\Gamma^{(n)} & \equiv & 2 (\mat X - \mat X_n)^t\mat g_n.
\eea
Equation~\ref{eq:appDE} is similar to \eq{eq:DE@CI} but can be used to approximate $\Omega^2$ about any nuclear geometry in the vicinity of a \gls{CI} (i.e. when $\norm{\mat X-\mat X_n}$ is sufficiently small and \eq{eq:ddO2} applies).

Imposing the constraint $\Omega=0$ is then equivalent to imposing the constraints $\Delta=\Gamma=0$.
We can now derive the procedure that we will use in \gls{ALM}.
Following the work of Yarkony and coworkers~\cite{Manaa:1993/jcp/5251}, we first define the following Lagrangian
\bea
\mathcal{L} & = & \Sigma + \lambda_1 \Delta + \lambda_2 \Gamma,
\eea
where $\lambda_1$ and $\lambda_2$ are Lagrange multipliers.
Applying the stationary condition with respect to all variables ($\mat X$, $\lambda_1$, and $\lambda_2$) results in a system of equations
\bea
\mat\nabla\mathcal{L} & = & \mat\nabla\Sigma + \lambda_1\mat\nabla\Delta + \lambda_2\mat\nabla\Gamma = \mat 0, \\
\frac{\partial\mathcal{L}}{\partial\lambda_1} & = & \Delta = 0, \\
\frac{\partial\mathcal{L}}{\partial\lambda_2} & = & \Gamma = 0.
\eea
To further simplify the notation, we define the new quantities:
\bea
\mat\varepsilon = (\begin{matrix}\Delta&\Gamma\end{matrix})^t, 
&\text{ and }&
\mat\lambda = (\begin{matrix}\lambda_1&\lambda_2\end{matrix})^t.
\eea
To apply the Newton algorithm, we first expand this set of coupled equations in the vicinity of $\mat X_n$ up to first order and utilize these approximate forms in the stationary conditions:
\bea
\mat\nabla\mathcal{L}^{(n)} & = & \mat s_n + \mat B_n\mat\lambda + \mat S_n (\mat X_{n+1} - \mat X_n) = \mat 0, \\
\mat\varepsilon^{(n)} & = & \mat\varepsilon_n + \mat B_n^t(\mat X_{n+1} - \mat X_n) = \mat 0.
\eea
where we defined $\mat s=\mat\nabla\Sigma$, $\mat S=\mat\nabla\mat\nabla^t\Sigma$, and $\mat X_{n+1}$ is an approximate position of the \gls{CI} in the vicinity of $\mat X_n$.
We can then solve this system of equations to obtain $\mat X_{n+1}$ as
\bea\label{eq:XALM}
\mat X_{n+1} & = & \mat X_n
- [ \mat 1_\mathcal{D} - \mat S_n^{-1} \mat B_n (\mat B_n^t\mat S_n^{-1}\mat B_n)^{-1} \mat B_n^t ] \mat S_n^{-1} \mat s_n 
\nonumber\\&&
- \mat S_n^{-1} \mat B_n(\mat B_n^t\mat S_n^{-1}\mat B_n)^{-1} \mat\varepsilon_n.
\eea
This procedure is then repeated iteratively until a \gls{MECI} is found, \ie we found a geometry $\mat X_f$ such that the quantities $\Omega_f$ and $\mat s_f^t(\mat 1_\mathcal{D}-\mat B_f (\mat B_f^t\mat B_f)^{-1} \mat B_f^t)\mat s_f$ are smaller than some threshold.
Since the Hessian $\mat\nabla\mat\nabla^t\Sigma=\mat S$ is not assumed to be known, we use the quasi-Newton algorithm to iteratively construct this Hessian along the optimization.
We note that the depicted algorithm differs from reference~\onlinecite{Manaa:1993/jcp/5251} in the absence of the Hessians of $\Delta$ and $\Gamma$ such that the Lagrange multipliers do not appear in the step calculation \eq{eq:XALM}.
This further alleviates us from the need of using extrapolatable functions~\cite{Yarkony:2004/jpca/3200} to ensure that the Lagrangian Hessian can be updated by the quasi-Newton algorithm.

To avoid the calculation of the \gls{DC}, $\mat g_n$, we use a fitting procedure based on \eq{eq:appDE}.
Since the energies and energy gradients are evaluated at each geometry along the optimization, we use the quantities $\Omega_{n-1}$, $\Omega_{n}$, $\mat d_{n-1}$, $\mat d_{n}$ calculated at consecutive geometries to extract $\mat g_n$.
A straightforward approach would be to solve \eq{eq:appDE} for $\mat g_n$.
This is possible if $\Omega_{n-1}^2 > [\Omega_n + (\mat X_{n-1} - \mat X_n)^t\mat d_n]^2$, and we get
\bea\label{eq:directsolve}
\mat g_n & = & \frac{ \Omega_{n-1}\mat d_{n-1} - [\Omega_n + (\mat X_{n-1} - \mat X_n)^t\mat d_n]\mat d_n }{\sqrt{ \Omega_{n-1}^2 - [\Omega_n + (\mat X_{n-1} - \mat X_n)^t\mat d_n]^2 }}.
\eea
Note that in \eq{eq:directsolve}, $\mat g_n$ is expressed as a linear combination of $\mat d_n$ and $\mat d_{n-1}$, and therefore, the \gls{BS} spans the two directions given by $\mat d_n$ and $\mat d_{n-1}$.
This observation agrees with the update scheme defining the \gls{UBS} method.~\cite{Maeda:2010/jctc/1538}
However, $\Omega_{n-1}^2 - [\Omega_n + (\mat X_{n-1} - \mat X_n)^t\mat d_n]^2$ may not be positive, so that \eq{eq:directsolve} is not valid anymore.
Hence, we prefer the following procedure, in which we only assume that the mathematical form of $\Omega$ is given by the model
\bea\label{eq:modDE}
\Omega_{M}(\mat X) & = & \sqrt{ [ c + \mat v^t(\mat X-\mat X_{n}) ]^2 + 4[ \mat w^t(\mat X-\mat X_{n}) ]^2 }, \nonumber\\
\eea
where $c$, $\mat v$, and $\mat w$ are unknown parameters.
For the model \eq{eq:modDE}, the \gls{DC} in $\mat X_n$ is given by $\mat g_n=\mat w$.
Fitting \eq{eq:modDE} is in fact equivalent to fitting and using the parameters of a \acrlong{LVC} model~\cite{Koppel:2004/175,Zhu:2009/jcp/234108,Zhu:2010/jcp/104101} at each step of the \gls{MECI} optimization.
For this purpose, we define the simple penalty vector function $\mat Z(\mat p)$ where $\mat Z$ is the ``error'' vector, and $\mat p$ is the vector of parameters:
\bea
\mat Z & = & \begin{pmatrix}\Omega_n-\Omega_M(\mat X_n),
\\\mat d_n-\mat\nabla\Omega_M|_{\mat X_n}
\\\Omega_n-\Omega_M(\mat X_{n-1}),
\\\mat d_{n-1}-\mat\nabla\Omega_M|_{\mat X_{n-1}}
\end{pmatrix}, \label{eq:defZ}\\
\mat p & = & (c,\mat v^t,\mat w^t)^t. \label{eq:defp}
\eea
We then apply an iterative Newton-related method on the equation $\mat Z=\mat 0$ to obtain $\mat w$, this approach is described in App.~\ref{app:fit}.
We then use $\mat w$ as an approximation for $\mat g_n$ in \eq{eq:XALM}.
This last paragraph is similar in spirit to previously designed methods for extracting the \gls{DC} from energy calculations evaluated at neighboring geometries~\cite{Koppel:2004/175,Kammeraad:2016/jpcl/5074,Gonon:2017/jcp/114114,Gonon:2017/jcp/114114,Baeck:2017/jcp/064107,Fdez.Galvan:2023/jctc/3418,Vandaele:2024/jctc/} and shares a lot in common with the particular method developed in Ref.~\onlinecite{Fdez.Galvan:2023/jctc/3418}, the main difference comes from the fact that we still employ the original \gls{LM} algorithm instead of employing the diabatic representation, and we do not utilize \gls{DC} in the \gls{MECI} optimization.

\subsection{The single-constraint method: \protect\gls{SLM}}

Observing that $\mat\nabla\Omega^2$ is well-defined at the \gls{CI} and that the Hessian $\mat\nabla\mat\nabla^t\Omega^2$ contains all the information on the \gls{BS}, prompted us to develop a method based on the single constraint $\Omega^2=0$.
The Lagrangian is given by
\bea
\mathcal{L} & = & \Sigma + \lambda\Omega^2.
\eea
The stationary condition is given by the set of coupled equations
\bea
\mat\nabla\mathcal{L} & = & \mat s + \lambda\mat\nabla\Omega^2, \\
\frac{\partial\mathcal{L}}{\partial\lambda} & = & \Omega^2 = 0.
\eea
Expanding these conditions to first order at $\{\mat X_n,\lambda_n\}$ and defining $\mat k=\mat\nabla\Omega^2=2\Omega\mat d$ and $\mat K=\mat\nabla\mat\nabla^t\Omega^2$, we obtain new relations involving the updated quantities $\mat X_{n+1}$ and $\lambda_{n+1}$
\bea
\mat\nabla\mathcal{L}^{(n)} & = & 
\mat s_n + \lambda_{n+1}\mat k_n + ( \mat S_n + \lambda_n\mat K_n ) (\mat X_{n+1}-\mat X_n) = \mat 0, \nonumber\\\\
\label{eq:cstrSLM}
(\Omega^{(n)})^2 & = & \Omega_n^2 + \mat k_n^t(\mat X-\mat X_n)= 0.
\eea
In order to retain information on the \gls{BS}, we must keep the term containing $\mat K_n$.
Hence, we must also update the Lagrange multiplier $\lambda$ in the \gls{SLM} method. 
The solution to this system of equations is 
\bea\label{eq:XSLM}
\mat X_{n+1} & = & \mat X_n - ( \mat S_n + \lambda_n\mat K_n )^{-1} ( \mat s_n + \lambda_{n+1}\mat k_n ), \\\label{eq:XSLM2}
\lambda_{n+1} & = & \frac{ \Omega_n^2 - \mat k_n^t( \mat S_n + \lambda_n\mat K_n )^{-1} \mat s_n }{ \mat k_n^t( \mat S_n + \lambda_n\mat K_n )^{-1} \mat k_n }.
\eea

The \gls{DC} is not needed since it does not appear in the formulation.
It is in fact ``replaced'' by the evaluation of the Hessian $\mat\nabla\mat\nabla^t\Omega^2=\mat K$.
Since we assume that Hessians are not calculated exactly, we utilize the quasi-Newton algorithm to iteratively construct both $\mat S$ and $\mat K$ along the optimization.

%%%%%%%%%%%%%%%%%%%%%%%%%%%%%%%%%%%%%%%%%%%%%%%%%%%%%%%%%%%%%%%%%%%%%%
\section{Computational details}
\label{sec:details}

\subsection{Algorithm for optimizations of \glspl{MECI}}

% Details on the Lagrange-Newton algorithm
Since we employ the quasi-Newton algorithm for the \gls{CI},
we use the \gls{BFGS} algorithm for the Hessians' update scheme.~\cite{Fletcher:2000/55}
The initial Hessians of $\Sigma$ are chosen to be $\mat S_0=0.5\times\mat 1_\mathcal{D}$ a.u. for all molecular systems but the benzene, for which $\mat S_0=\mat 1_\mathcal{D}$ a.u.
Regarding the \gls{SLM} method, the initial Hessians of $\Omega^2$ are chosen to be $\mat K_0=\mat 0$ a.u.
In \gls{SLM}, we also need to optimize the Lagrange multiplier $\lambda$.
The initial value for $\lambda$ is chosen to be $\lambda_0=0.1$ a.u.
However, the initial value does not have any impact on the convergence since it multiplies $\mat K_0=\mat 0$ (see Eqs.~(\ref{eq:XSLM}-\ref{eq:XSLM2}) for $n=0$).

% Convergence criteria
Two convergence criteria were chosen to test if the current geometry is a \gls{MECI}.
The first criterion estimates how close is the geometry to the \gls{IS}, which must be such that $\Omega<5\times 10^{-4}$ a.u.
The second criterion estimates how close is the current geometry to a minimum of the \gls{IS}.
For this purpose, we evaluate the \gls{RMS} of the projected gradient and convergence is assumed if $\sqrt{\mat s^t\mat P_{IS}\mat s/\mathcal{D}}<5\times 10^{-4}$ a.u.
Since the \gls{DC} is not known, $\mat P_{IS}$ is also \emph{a priori} unknown.
However, we can utilize the observation, made from \eq{eq:directsolve}, that \gls{DC} can be approximated by a linear combination of $\mat d$ and $\mat d_{n-1}$.
Hence, the \gls{BS} is approximated as the $2$-dimensional subspace spanned by $\mat d_n$ and $\mat d_{n-1}$:
\bea
\mat P_{BS} & \approx & (\begin{matrix} \mat d_n & \mat d_{n-1} \end{matrix}) 
[ (\begin{matrix} \mat d_n & \mat d_{n-1} \end{matrix})^t (\begin{matrix} \mat d_n & \mat d_{n-1} \end{matrix}) ]^{-1} 
(\begin{matrix} \mat d_n & \mat d_{n-1} \end{matrix})^t. \label{eq:approxPBS} \nonumber\\
\eea
This approximation is similar in spirit to the approximate scheme employed in the \gls{UBS} method.~\cite{Maeda:2010/jctc/1538}

% Linesearch variation
When the quasi-Newton step gives a new geometry that is further away from the \gls{MECI} than the current geometry (when $\Omega$ or $\Sigma$ increases), we use a linesearch~\cite{Schlegel:2011/wcms/790} to iteratively try reducing the step size.
We use a very basic implementation where the step is simply divided by $2$ for a maximum of $5$ "linesearch iterations".
To reduce the number of electronic structure calculation, we added the following conditions: i) we impose a maximum step size of $0.2$ a.u., ii) we accept small increase of $\Omega$ and $\Sigma$ scaled by the changes of the previous step, \ie if $(\Sigma_n-\Sigma_{n-1})<50|\Sigma_{n-1}-\Sigma_{n-2}|$ and $(\Omega_n-\Omega_{n-1})<10|\Omega_{n-1}-\Omega_{n-2}|$, and iii) the linesearch is limited to five iterations maximum (down to $3$ \% of the maximum step).

The methods were implemented in a Fortran 90 code using the LAPACK library.~\cite{LAPACK} 
This code is then interfaced with electronic structure calculations packages used for the adiabatic energies and gradients: the Gaussian g16 package,~\cite{g16} and the Molpro package.~\cite{Werner:2020/jcp/144107}

% Other methods
We also implemented the \gls{LM}, \gls{CG}, and \gls{UBS} methods for comparison purpose.
Our implementation of the \gls{LM} method~\cite{Manaa:1993/jcp/5251} is simply the \gls{ALM} method for which we use the exact \gls{DC} calculated by the electronic structure package, and the \gls{BS} projector is calculated exactly using Eqs.~(\ref{eq:Pbs}-\ref{eq:B}).
It should be noted that, as opposed to the reference \onlinecite{Yarkony:2004/jpca/3200}, we do not use second order information on $\Delta$ and $\Gamma$.
The \gls{CG} method~\cite{Bearpark:1994/cpl/269} is implemented using the following definition:
\bea\label{eq:gc}
\mat g_{cg} & = & c_{cg}'[(1-c_{cg})\mat P_{IS}\mat s + 2c_{cg}{\Omega\mat d}/{\norm{\mat d}}],
\eea 
where we chose $c_{cg}=0.9$ according to reference~\onlinecite{Keal:2007/tca/837} 
and found that a scaling $c_{cg}'=0.2$ was optimal for our test group of molecules.
We solve the equation $\mat g_{cg}=\mat 0$ using the quasi-Newton/linesearch algorithm describe in this section, with the initial Hessian being initialized as $(1-c_{cg})\mat S_0/2+c_{cg}{\mat d\mat d^t}/{\norm{\mat d}}$.
As for the \gls{LM} method, the \gls{BS} projector is calculated exactly using Eqs.~(\ref{eq:Pbs}-\ref{eq:B}), and $\mat P_{IS}=\mat 1_\mathcal{D}-\mat P_{BS}$ in \eq{eq:gc}.
Our implementation of the \gls{UBS} method follows our implementation of the \gls{CG}, except that $\mat P_{BS}$ is calculated approximately utilizing the prescription from reference~\onlinecite{Maeda:2010/jctc/1538}: 
i) the initial projector is calculated as
\bea
\mat P_{BS}(\mat X_0) & \approx & (\begin{matrix} \mat d_0 & \mat s_0 \end{matrix}) 
[ (\begin{matrix} \mat d_0 & \mat s_0 \end{matrix})^t (\begin{matrix} \mat d_0 & \mat s_0 \end{matrix}) ]^{-1} 
(\begin{matrix} \mat d_0 & \mat s_0 \end{matrix})^t, \nonumber\\
\eea
and ii) the projector is updated following the construction (equivalent to Ref.~\onlinecite{Maeda:2010/jctc/1538})
\bea
\mat P_{BS}(\mat X_{n+1}) 
& = & \mat P_{BS}(\mat X_{n}) + \frac{\mat d_{n+1}\mat d_{n+1}^t}{\mat d_{n+1}^t\mat d_{n+1}} \nonumber\\
&&- \frac{\mat P_{BS}(\mat X_n)\mat d_{n+1}\mat d_{n+1}^t\mat P_{BS}(\mat X_n)}{ \mat d_{n+1}^t \mat P_{BS}(\mat X_n)\mat d_{n+1} }.
\eea

\subsection{Electronic structure calculations}

For the tests calculations on the small molecules (ethylene, benzene, diazomethane, and methaniminium), we optimize the \gls{MECI} between the S$_0$ and $S_1$ surfaces.
We use the \gls{CASSCF} method with orbital state-averaging between the intersecting states.
We chose a small basis set, STO-3G, for the test-molecules since the aim is to test and compare numerical methods rather than providing results of chemical accuracy.
The active spaces employed in each case are described in Tab~\ref{tab:CASs}, and have been taken from the literature,~\cite{Barbatti:2004/jcp/2004,Palmer:1993/jacs/673,Yamamoto:1994/jacs/2064,Page:2003/jcc/298} except for the methaniminium system for which we chose the minimal active space of two electrons in two orbitals.
The calculations were done using the Gaussian g16 package.~\cite{g16}

\begin{table}[!bht]
  \centering
  \caption{Definition of the active space used in the \gls{CASSCF} method for each test molecule with the format (number of active electrons, number of active orbitals).}
    \newcommand{\size}{0.1cm}
    \begin{tabular}{@{\hspace{\size}}l@{\hspace{\size}}c@{\hspace{\size}}c@{\hspace{\size}}c@{\hspace{\size}}c@{\hspace{\size}}}
    \hline
Molecules & Ethylene & Benzene & Methaniminium & Diazomethane \\\hline
CAS       & (2,2)~\cite{Barbatti:2004/jcp/2004}    & (6,6)~\cite{Palmer:1993/jacs/673}   & (2,2)        & (6,6)~\cite{Yamamoto:1994/jacs/2064,Page:2003/jcc/298}        \\
    \hline
    \end{tabular}
  \label{tab:CASs}
\end{table}

Following Ref.~\onlinecite{Sahihi:2020/jpcb/2539}, 
the furimamide \gls{MECI} of interest is between S$_1$ and S$_2$. 
We employed the \gls{TDDFT} with the CAM-B3LYP~\cite{Yanai:2004/cpl/51} functional together with Pople’s polarization valence-triple-$\zeta$ 6-311G-(2d,p) basis set.~\cite{Hehre:1972/jcp/2257,Hariharan:1973tca/213,Francl:1982/jcp/77}
Furimamide's calculations were done using the Gaussian g16 package.~\cite{g16}

Regarding the S$_0$/S$_2$ \gls{MECI} of the silver trimer, calculations were done following Ref.~\onlinecite{Mitrushchenkov:2023/pcp/e202300317}.
We employed the \gls{MSCASPT2} level theory with the augmented polarized correlation-consistent triple-$\zeta$ aug-cc-pVTZ-PP basis set~\cite{Figgen:2005/cp/227} including a small (28-core-electron) relativistic pseudo-potential.
We employed a minimal (3,5) active space for state-average calculations including the two lowest electronic states, and a level-shift 0f 0.2 in the CASPT2 step~\cite{Celani:2000/jcp/5546,Roos:1995/cpl/215} with 12 core orbitals (uncorrelated 4s and 4p orbitals).
Calculations for Ag$_3$ were all done using the Molpro package.~\cite{Werner:2020/jcp/144107}

%%%%%%%%%%%%%%%%%%%%%%%%%%%%%%%%%%%%%%%%%%%%%%%%%%%%%%%%%%%%%%%%%%%%%%
\section{Test of the methods}
\label{sec:test}

Both the \gls{ALM} and \gls{SLM} methods are tested on four different small molecular systems for which we can easily obtain the \gls{MECI} by employing the usual optimization algorithm, namely \gls{LM} and \gls{CG}.~\cite{Keal:2007/tca/837}
These molecular systems are: ethylene,~\cite{Barbatti:2004/jcp/2004} benzene,~\cite{Palmer:1993/jacs/673} methaniminium,~\cite{Barbatti:2006/mp/1053} and diazomethane.~\cite{Yamamoto:1994/jacs/2064,Page:2003/jcc/298}

Initial geometries are constructed as follows for each molecular systems.
For ethylene, the molecule is first optimized to a minimum in the ground electronic state.
The molecule was subsequently distorted by setting the dihedral angle around the double bond to 90 degrees.
The initial geometry construction for the methaniminium followed the same scheme as for ethylene, except that the electronic states under consideration are cationic states.
For diazomethane, the initial geometry is simply taken as the ground electronic state minimum.
Regarding benzene, we optimized the geometry to the minimum of the ground electronic state and the brought one of the CH out of the molecular plane by $0.5$ \AA.

\begin{table*}[!bht]
  \centering
  \caption{Comparison of the convergence properties between the different methods.
            Energies are given in Hartree while \glspl{RMSD} are given in \AA.
            The \gls{RMSD} is calculated as follows:
            RMSD$(\mat X)=\sqrt{ \norm{\mat X_0-\mat Q}^2/N_\text{at} }$,
            where $\mat X_0$ is some reference geometry from which we want to compute the RSMD with $\mat X$,
            $\mat Q=\mat R(\mat X-\mat X_{CM})+\mat X_0^{CM}$ is the geometry $\mat X$ but translated to the mass center $\mat X_0^{CM}$ and rotated by the orthogonal transformation $\mat R$ to align both geometries by minimizing the \gls{RMSD},~\cite{Wahba:2006/sr/409} 
            and $N_\text{at}$ is the number of atoms.
            RMSD/LM indicates that the reference $\mat X_{CI}$ is obtained by the \gls{LM} method (\gls{ALM} for furimamide and Ag$_3$), while RMSD/CG indicates that $\mat X_{CI}$ is obtained by the \gls{CG} method (\gls{UBS} for furimamide and Ag$_3$).}
    \newcommand{\size}{0.25cm}
    \begin{tabular}{@{\hspace{\size}}l@{\hspace{\size}}c@{\hspace{\size}}c@{\hspace{\size}}c@{\hspace{\size}}c@{\hspace{\size}}c@{\hspace{\size}}c@{\hspace{\size}}}
    \hline
Molecule      & Method & Steps &  $\Sigma/2$ &     $\Omega$      &      RMSD/LM          (RMSD/CG)         \\
              &        &       &             &                   & (*)  RMSD/ALM         (RMSD/UBS)        \\\hline
Ethylene      &   LM   & $ 20$ & $ -76.8370$ & $3.7\cdot10^{-7}$ &                     \\
              &   CG   & $ 51$ & $ -76.8370$ & $3.9\cdot10^{-5}$ &  $5.2\cdot10^{-1}$  \\
              &  UBS   & $ 64$ & $ -76.8370$ & $9.5\cdot10^{-5}$ &  $5.2\cdot10^{-1}$   $( 1.9\cdot10^{-3})$\\
              &  ALM   & $ 27$ & $ -76.8370$ & $2.6\cdot10^{-4}$ &  $2.6\cdot10^{-3}$  \\
              &  SLM   & $ 26$ & $ -76.8371$ & $1.2\cdot10^{-4}$ &  $1.0\cdot10^{-3}$  \\\hline
Benzene       &   LM   & $ 27$ & $-227.7809$ & $5.0\cdot10^{-7}$ &                     \\
              &   CG   & $ 62$ & $-227.7809$ & $2.8\cdot10^{-5}$ &  $2.6\cdot10^{-3}$  \\
              &  UBS   & $ 69$ & $-227.7809$ & $1.7\cdot10^{-5}$ &  $1.8\cdot10^{-3}$   $( 1.6\cdot10^{-3})$\\
              &  ALM   & $ 36$ & $-227.7809$ & $5.7\cdot10^{-6}$ &  $2.0\cdot10^{-3}$  \\
              &  SLM   & $ 28$ & $-227.7809$ & $4.5\cdot10^{-5}$ &  $2.5\cdot10^{-3}$  \\\hline
Diazomethane  &   LM   & $ 17$ & $-145.9964$ & $2.8\cdot10^{-6}$ &                     \\
              &   CG   & $ 39$ & $-145.9964$ & $1.9\cdot10^{-5}$ &  $5.8\cdot10^{-1}$  \\
              &  UBS   & $ 37$ & $-145.9957$ & $9.7\cdot10^{-5}$ &  $1.4\cdot10^{-2}$   $( 5.9\cdot10^{-1})$\\
              &  ALM   & $ 23$ & $-145.9964$ & $2.6\cdot10^{-5}$ &  $5.8\cdot10^{-1}$  \\
              &  SLM   & $ 34$ & $-145.9965$ & $2.8\cdot10^{-4}$ &  $5.8\cdot10^{-1}$  \\\hline
Methaniminium &   LM   & $ 16$ & $ -93.0916$ & $4.7\cdot10^{-7}$ &                     \\
              &   CG   & $ 29$ & $ -93.0916$ & $5.5\cdot10^{-5}$ &  $2.4\cdot10^{-3}$  \\
              &  UBS   & $ 41$ & $ -93.0916$ & $2.7\cdot10^{-5}$ &  $1.3\cdot10^{-3}$   $( 3.4\cdot10^{-3})$\\
              &  ALM   & $ 20$ & $ -93.0916$ & $2.0\cdot10^{-5}$ &  $1.1\cdot10^{-3}$  \\
              &  SLM   & $ 42$ & $ -93.0916$ & $1.7\cdot10^{-4}$ &  $4.1\cdot10^{-3}$  \\\hline
Furimamide    &  ALM   & $ 36$ &$-1202.1388$ & $1.4\cdot10^{-6}$ &                      \\
              &  UBS   & $125$ &$-1202.1389$ & $8.7\cdot10^{-5}$ &  $2.7\cdot10^{-2}$(*)\\
              &  SLM   & $ 69$ &$-1202.1392$ & $1.5\cdot10^{-5}$ &  $5.8\cdot10^{-2}$ $( 3.6\cdot10^{-2} )$(*)\\\hline
Ag$_3$        &  ALM   & $  8$ &$-439.5067$ & $1.5\cdot10^{-4}$ &                     \\
              &  UBS   & $  7$ &$-439.5067$ & $6.4\cdot10^{-6}$ &  $1.5\cdot10^{-3}$(*)\\
              &  SLM   & $  5$ &$-439.5067$ & $5.1\cdot10^{-6}$ &  $2.0\cdot10^{-3}$ $( 7.3\cdot10^{-4})$(*)\\
    \hline
                  \end{tabular}
  \label{tab:CASs}
\end{table*}

\begin{figure}
  \centering
  \includegraphics[width=0.45\textwidth]{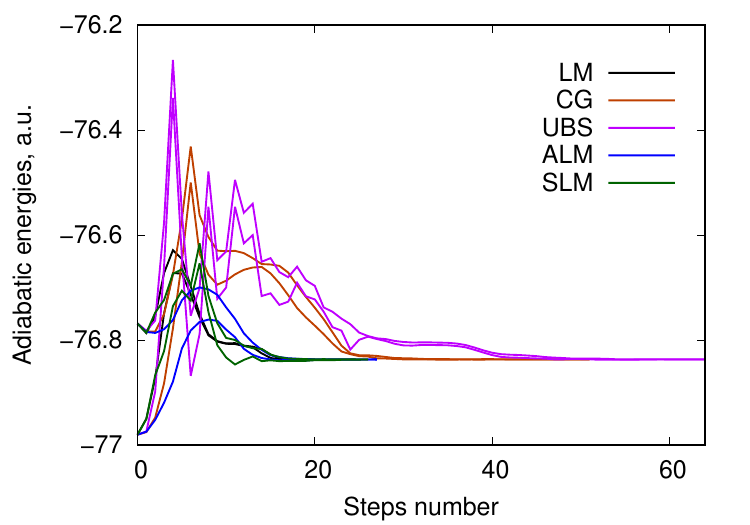}
  \caption{Energies of S$_0$ and S$_1$ states of the ethylene molecule along the \gls{MECI} optimization.}
  \label{fig:ethylene-energies}
\end{figure}

\begin{figure}
  \centering
  \includegraphics[width=0.45\textwidth]{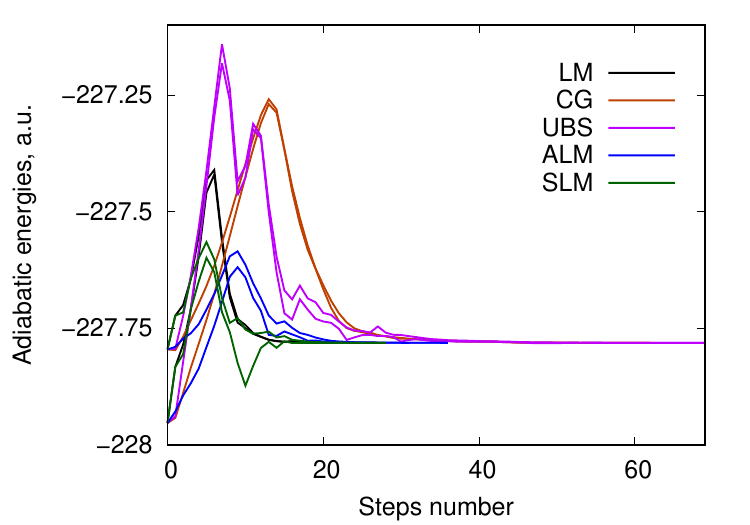}
  \caption{Energies of S$_0$ and S$_1$ states of the benzene molecule along the \gls{MECI} optimization.}
  \label{fig:benzene-energies}
\end{figure}

\begin{figure}
  \centering
  \includegraphics[width=0.45\textwidth]{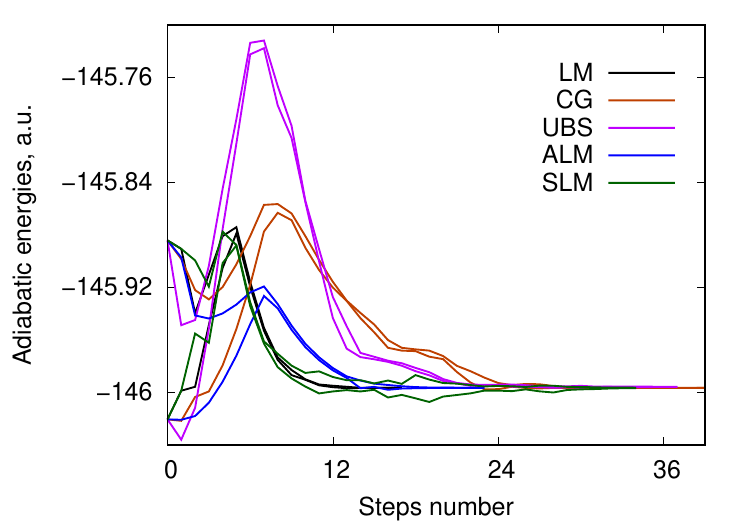}
  \caption{Energies of S$_0$ and S$_1$ states of the diazomethane molecule along the \gls{MECI} optimization.}
  \label{fig:diazomethane-energies}
\end{figure}

\begin{figure}
  \centering
  \includegraphics[width=0.45\textwidth]{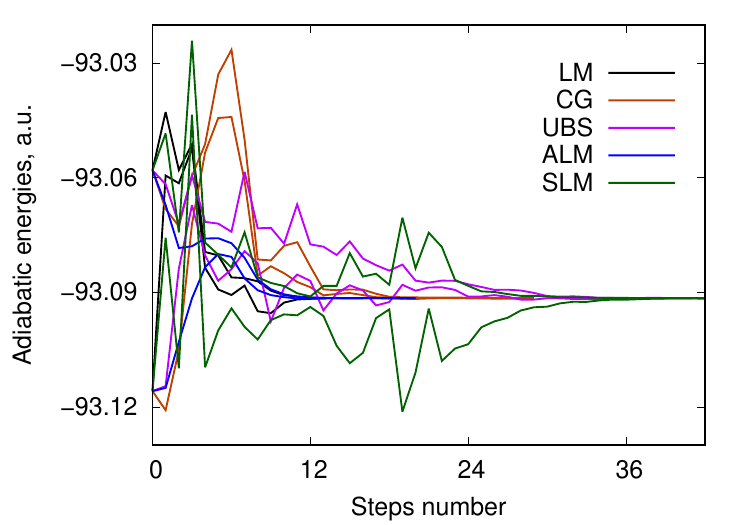}
  \caption{Energies of S$_0$ and S$_1$ states of the methaniminium molecule along the \gls{MECI} optimization.}
  \label{fig:methaniminium-energies}
\end{figure}

The results of the optimizations for each molecular system with each method are given in Tab.~\ref{tab:CASs} and the energies' evolution for each optimization is given in Figs.~(\ref{fig:ethylene-energies}-\ref{fig:methaniminium-energies}).
Results from Tab.~\ref{tab:CASs} show that all methods converge qualitatively to the same \gls{MECI} geometries.
Indeed, for each molecular system, the average energies $\Sigma/2$ differ by a maximum of $8\cdot10^{-4}$ a.u. and the \glspl{RMSD} by $0.6$ \AA{}. 
For the tests involving benzene and methaniminium, the \gls{RMSD} is even below $5\cdot10^{-3}$ \AA{}.
In the ethylene case, we observe the \gls{RMSD} being $2.6\cdot10^{-3}$ at most \AA{} if we compare \gls{MECI} geometries obtained with similar methods, i.e. \gls{CG} compared to \gls{UBS} and \gls{ALM} and \gls{SLM} compared to \gls{LM}.
This shows that \gls{LM}-based methods converge to a slightly different \gls{MECI} geometry than \gls{CG}-based methods for ethylene.
The diazomethane's case shows a less homogeneous convergence if we compare \glspl{RMSD}.
However, the average energies $\Sigma/2$ obtained for each geometry are very close in energy, which indicates that the projected Hessian in the \gls{IS} contains flat modes (i.e three eigenvalues of the intersection space Hessian are lower than $4\cdot10^{-10}$ a.u. in absolute value while others are larger than $10^{-3}$ a.u.).

The number of steps necessary for convergence is always the lowest for the \gls{LM}.
By comparison, the number of steps necessary to converge with the \gls{CG} method is larger, often by more than twice, with an average ratio of $2.4$.
This observation is in agreement with Ref.~\onlinecite{Keal:2007/tca/837}.
We observe that the \gls{UBS} method follows the same trend, with an average ratio of $2.6$, which is to be expected since it is based on the \gls{CG} method.
Regarding the \gls{ALM} and \gls{SLM} methods, they perform better with average ratios of $1.3$ and $1.7$ respectively.
We must emphasize that the different methods were all implemented using the quasi-Newton method combined with a linesearch.
Our implementation is straightforward and might not be as effective for all methods.
In fact the tendencies that \gls{CG}-based methods require more steps on average to converge, might very well be explained by the fact that they are not based on the minimization of a scalar function, for which the quasi-Newton approach is designed.~\cite{Keal:2007/tca/837,Toniolo:2002/jpca/4679,Ruiz-Barragan/2013/jctc/1433}

The evolution of the S$_0$ and $S_1$ energies along the optimization steps for each method is given in Figs.~\ref{fig:ethylene-energies}-\ref{fig:methaniminium-energies}.
In agreement with the previous paragraph, these figures show numerical evidence that the \gls{LM}-based and the \gls{SLM} methods almost always reach the \gls{IS} space much before the \gls{CG}-based methods.
The \gls{LM} and \gls{SLM} methods show an analogous pattern for the first few steps, with an initial increase of both energies to a similar geometry in the \gls{IS}.
However, passed these first few steps, the \gls{LM} method keeps the subsequent geometries within the \gls{IS} while the \gls{SLM} method often deviates from it.
This observation is particularly severe for methaniminium (see Fig.~\ref{fig:methaniminium-energies}).
It indicates a difficulty of \gls{SLM} to properly define the \gls{IS} in order to optimize within this subspace.
Two numerical ingredients are suspected to cause these deviations: i) the approximate \gls{BFGS} construction of the Hessian which is supposed to contain all the information about the \gls{BS}, and 2) the fact that the constraint appearing in \eq{eq:cstrSLM} is trivially satisfied in the \gls{IS} since we have that $\Omega_n=0$ and $\mat k_n=\mat 0$, such that maintaining the constraint has less weight in the optimization step once the geometry is within the \gls{IS}.
The same figures show that the \gls{ALM} typically takes more steps to reach the \gls{IS}, but once a geometry is found within this subspace, the constraints properly impose the degeneracy for the remaining steps.
Hence, the \gls{ALM} method is overall more stable and converges faster than \gls{SLM} while the \gls{SLM} method seems to behave better far from the \gls{IS}.

We also wish to note that the approximate construction of the projector on the branching space (see Eq.~\ref{eq:approxPBS}), is successfully applied to evaluate the norm of the energy gradient within the intersection space.
This approximation provides a good convergence criteria for \gls{MECI} optimization at low computational cost.

%%%%%%%%%%%%%%%%%%%%%%%%%%%%%%%%%%%%%%%%%%%%%%%%%%%%%%%%%%%%%%%%%%%%%%
\section{Applications to more complex systems}
\label{sec:application}

In this section, we apply the developed methods to other molecular systems, namely the furimamide and the Ag3, for which the MECI needs to be characterized.
The \gls{DC} calculations are either unavailable or too costly in the electronic structure method of choice for these systems.

\subsection{Optimization of the furimamide's S$_1$/S$_2$ \gls{MECI}}

Furimamide is a bioluminescent molecule that attracted a lot of attention during the last ten years, since it became commercially available, and because it is two orders of magnitude brighter than the more commonly studied oxidized luciferins,~\cite{Hall:2012/acb/1848} with many applications.~\cite{Verhoef:2016/bba/284,Lackner:2015/nc/2041,Germain-Genevois:2016/mib/62,Shramova:2016/an/118,Shramova:2017/dbb/228}

\begin{figure}
  \centering
  \includegraphics[width=0.5\textwidth]{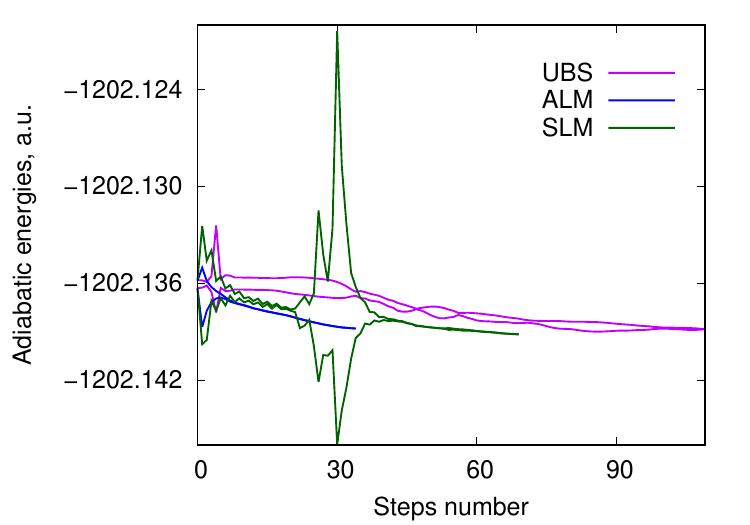}
  \caption{Energies of S$_1$ and S$_2$ states of the furimamide molecule along the \gls{MECI} optimization.}
  \label{fig:furimamide-energies}
\end{figure}

A recent study confirmed the existence of a crossing point between the first two excited electronic states~\cite{Sahihi:2020/jpcb/2539} that is key in the furimamide's de-excitation process.
Using the geometry of the CI found in reference~\onlinecite{Sahihi:2020/jpcb/2539} 
as an initial structure, we optimized the \gls{MECI} using the \gls{UBS}, \gls{ALM}, and \gls{SLM} methods. 
All three methods converged to the \gls{MECI}.
Results are presented in Tab.~\ref{tab:CASs} and in Fig.~\ref{fig:furimamide-energies}.

The Lagrange multipliers based methods, \gls{ALM} and \gls{SLM}, perform better than the \gls{UBS} method.
This is to be expected from the better performance of \gls{ALM} and \gls{SLM} regarding the tests on the previous set of molecules (in Sec.~\ref{sec:test}).
Indeed, the \gls{SLM} method converges almost twice faster than \gls{UBS}, and the \gls{ALM} method is almost four times faster.
Accounting for the fact that the energy difference is very small already at the very first geometry, a \acrlong{LVC} model for the \gls{PES} is valid.
Hence, the approximate calculation of the \gls{DC} in the \gls{ALM} method is already very efficient in the first steps of the optimization, which gives a certain advantage to this method.

\begin{figure}
  \centering
  \begin{tabular}{lc}
    a)\vspace{-1cm}&\\
    &\hspace{-0.1cm}\includegraphics[width=0.35\textwidth]{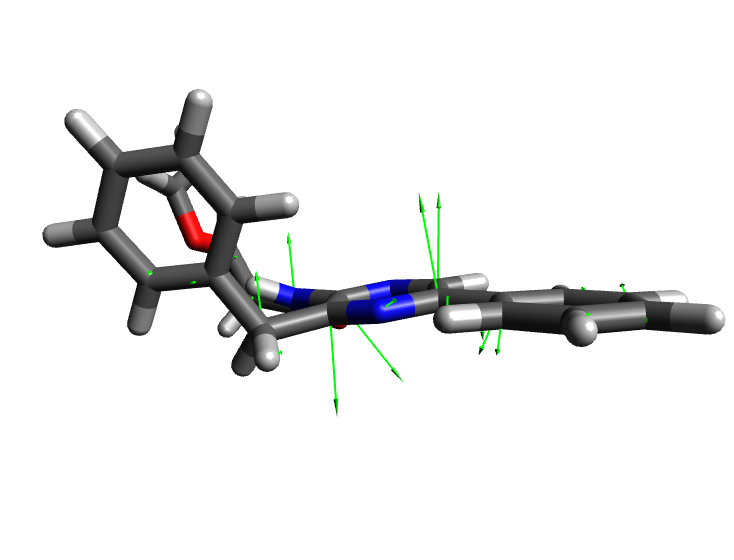}\\[-0.8cm]
    b)\vspace{-1cm}&\\
    &\hspace{-0.1cm}\includegraphics[width=0.35\textwidth]{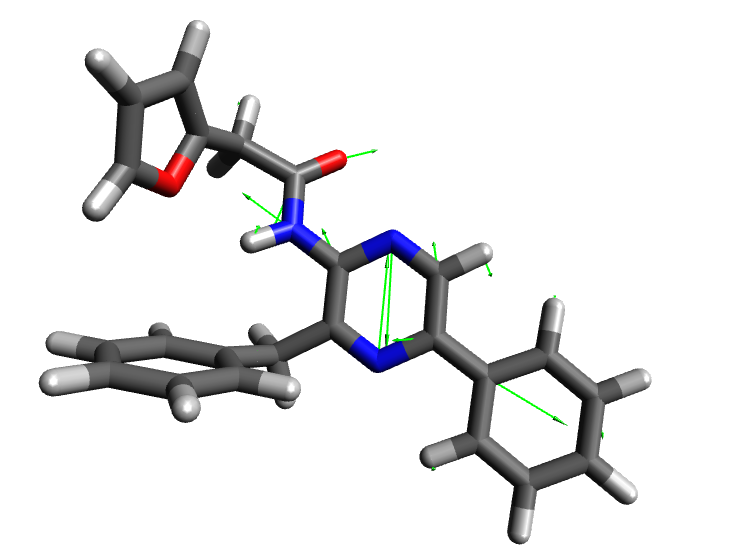}
  \end{tabular}
  \caption{Vectors of collective nuclear motions depicting two orthogonal directions of the branching space: a) out-of-plane motion of the central pyrazine cycle, b) bond length alternation of the pi system of the pyrazine cycle.}
\label{fig:BSvect}
\end{figure}

\begin{figure}
  \centering
  \begin{tabular}{ccl}
    \includegraphics[width=0.25\textwidth]{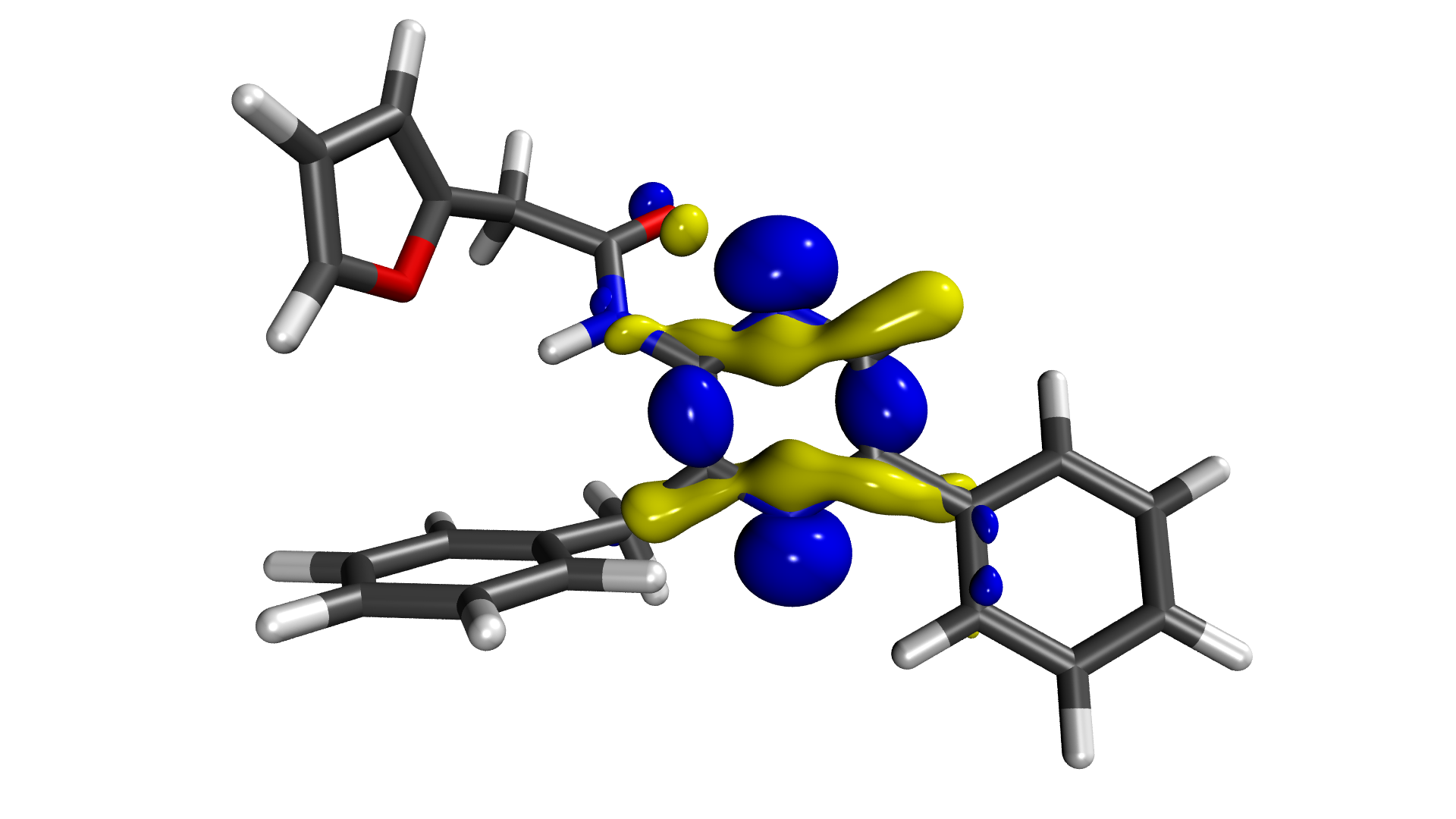}&\includegraphics[width=0.25\textwidth]{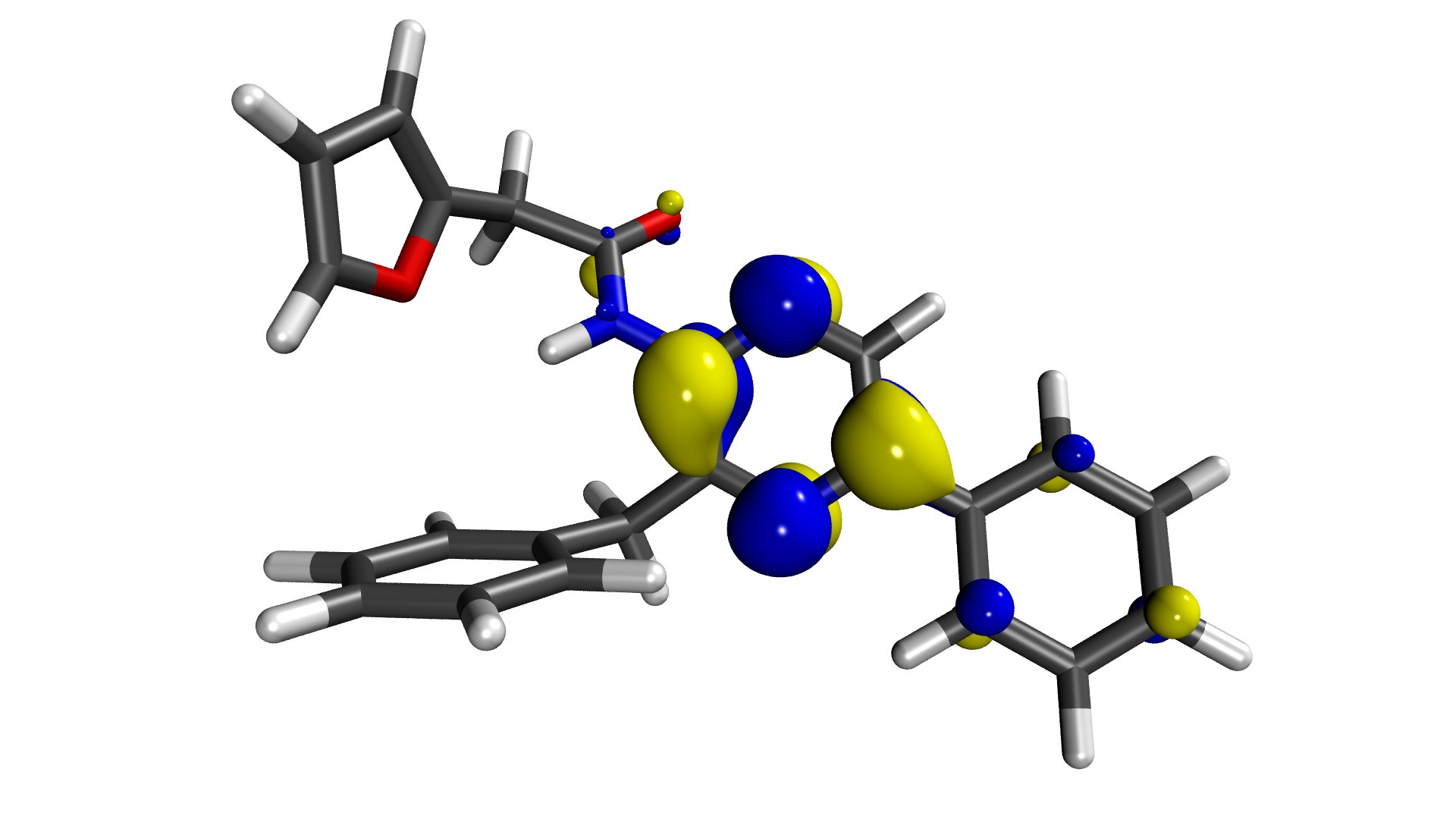}&\\
    n$_\text{N}$: S$_1$ hole &$\pi^*$: S$_1$ particle&\\
    \includegraphics[width=0.25\textwidth]{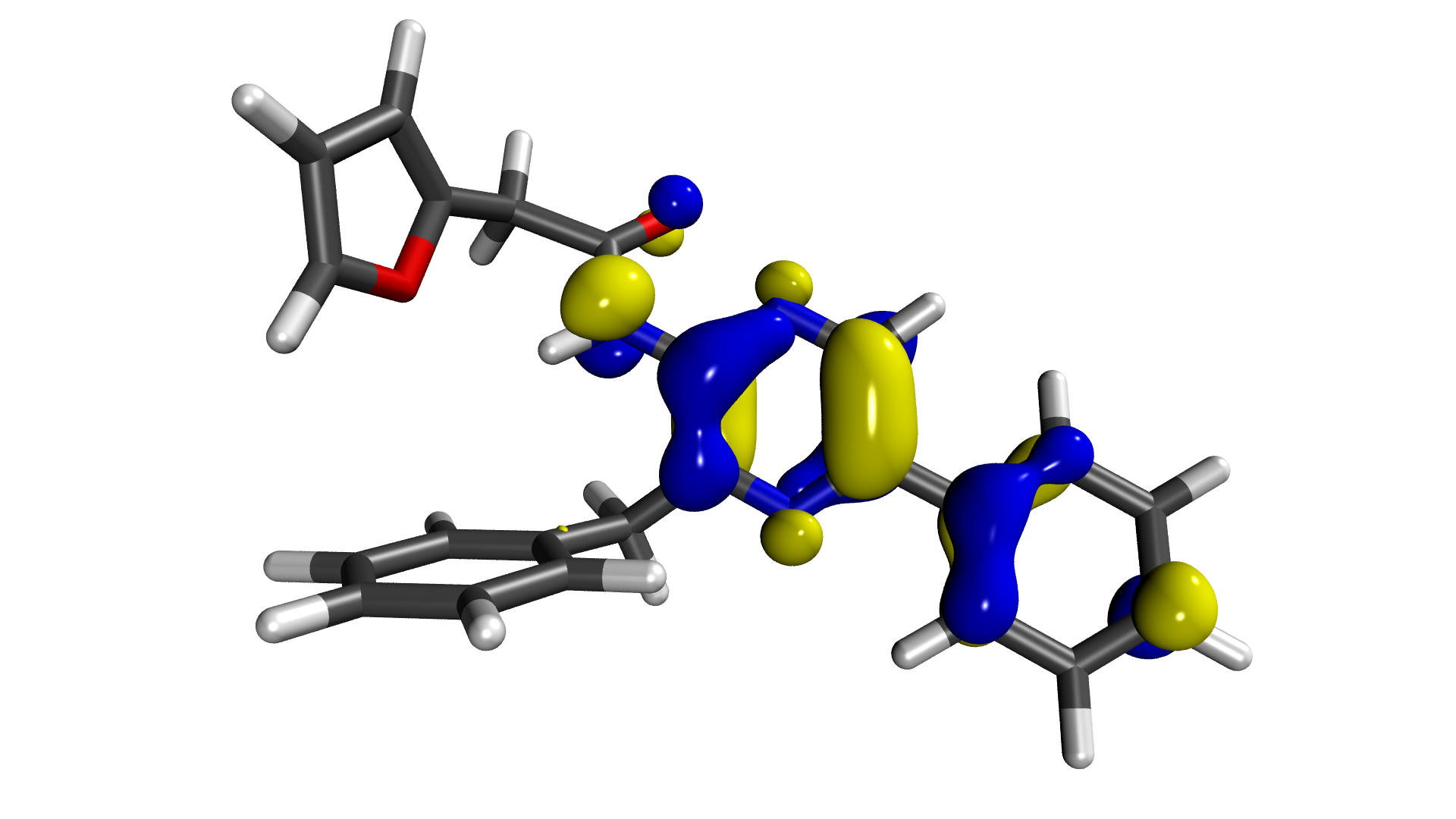}&\includegraphics[width=0.25\textwidth]{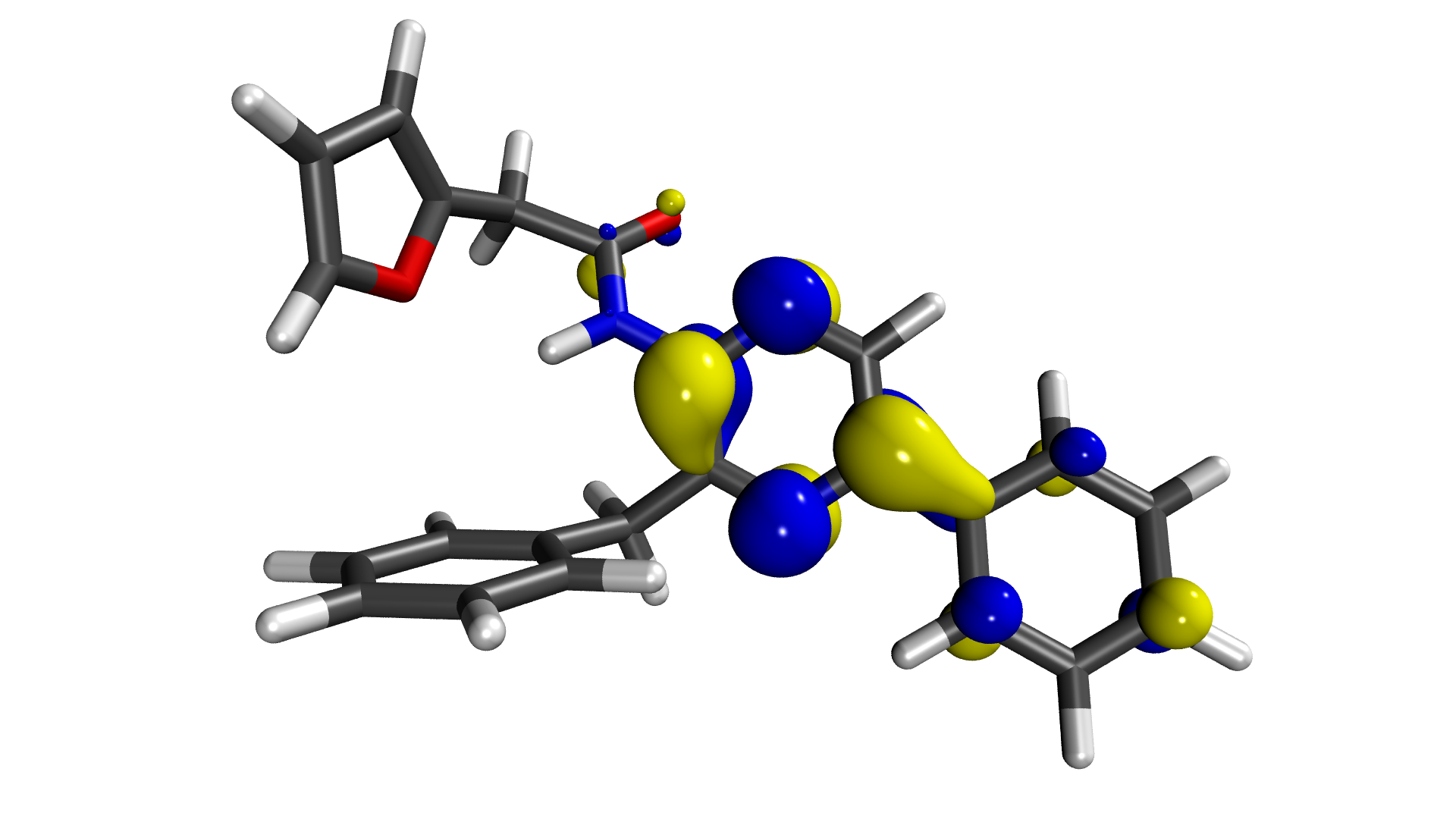}&\\
    $\pi$: S$_2$ hole&$\pi^*$: S$_2$ particle&\\
  \end{tabular}
  \caption{Description of the natural transition orbitals involved of the S$_1$ and S$_2$ electronic excited states that intersect at the \protect\gls{MECI} of the furimamide. 
           The orbitals are obtained at the \protect\acrlong{FC} geometry.}
  \label{fig:orb}
\end{figure}

We wish now to characterize the \gls{MECI} by studying the branching space vector.
Since, the energy difference at convergence is almost zero, \eq{eq:ddO2} applies.
According to this equation, the two eigenvectors with the largest eigenvalues span the \gls{BS}.
Thus, we use these two eigenvectors (already orthonormalized) as the \gls{BS} basis vectors.
This is in line with the constructions suggested in Refs.~\onlinecite{Kammeraad:2016/jpcl/5074,Gonon:2017/jcp/114114,Baeck:2017/jcp/064107,Vandaele:2024/jctc/}.
The first vector, depicted in Fig~\ref{fig:BSvect}-a, clearly shows an out-of-plane motion of the central pyrazine cycle.
This indicates pyramidalization of the corresponding carbon atoms, which, in turn, indicates that the $\pi$ electronic system of the pyrazine cycle has been weakened.
The second vector, presented in Fig~\ref{fig:BSvect}-b, depicts nuclear motions that are confined in the plane of the pyrazine, which indicate a restructuration of the $\pi$ system. 
This is in agreement with the electronic structure of exited states S1 and S2 as described in Ref.~\onlinecite{Sahihi:2020/jpcb/2539}. 
Indeed, both states are transitions mainly centered in the pyrazine moiety. 
The brightest state (S$_2$) at the Franck-Condon geometry is described as a $\pi^*\leftarrow\pi$, while the S$_1$ corresponds to a $\pi^*\leftarrow n_N$ transition, both centered on the pirazine aromatic ring. 
The natural transition orbitals~\cite{Martin:2003/jcp/4775} of both states at the \acrlong{FC} geometry is depicted in Fig.~\ref{fig:orb}. 
In agreement with the observation made in ref.~\onlinecite{Campetella:2020/jct/1156}, these natural transition orbitals describe the electronic states involved in the crossing in a compact manner.
This is also in line with the geometry distortions found for the simpler pyrazine molecule when the S$_2$/S$_1$ \gls{MECI} is reached.~\cite{Woywod:1994/jcp/1400,Taylor:2023/jcp/214115} 
This geometry distortions are described by not only the simultaneous elongation of C-N and C-C bonds but also the overall stretching of the aromatic ring along the axis bisecting the two nitrogen atoms.

\subsection{Optimization of the S$_0$/S$_1$ \gls{MECI} for Ag$_3$}

Recent advances in selective synthesis technique enable the production of atomically precise monodisperse metal clusters.~\cite{Lu:2012/csr/3594,Du:2020/cr/526} 
These clusters exhibit molecule-like properties that are distinct from the corresponding nanoparticle or bulk material.
In particular, these clusters commonly express a fluxional character related to a Jahn-Teller effect, that is of interest for various applications including for example catalysis or sensing.~\cite{deLara-Castells:2023/pccp/15081}
Such clusters present numerous challenges in their modelling due to their high number of electrons and their high symmetry.
Hence, electronic structure calculation commonly requires static correlation combined with utilizing pseudopotentials, in post-\gls{CASSCF} methods.
A recent case study of such a cluster is for example given in Ref.~\onlinecite{Mitrushchenkov:2023/pcp/e202300317} 
for the case of Cu$_5$, where \gls{MSCASPT2} calculations were employed.

\begin{figure}
  \centering
  \includegraphics[width=0.5\textwidth]{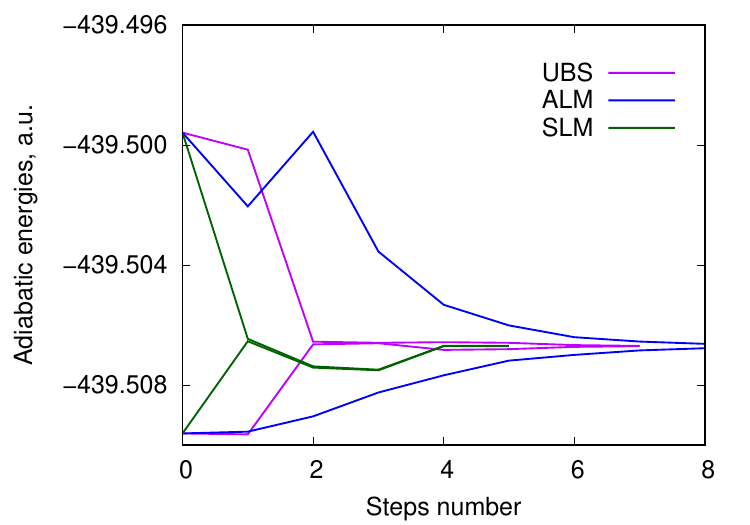}
  \caption{Energies of S$_0$ and S$_1$ states of Ag$_3$ along the \gls{MECI} optimization.}
  \label{fig:Ag3-energies}
\end{figure}

As a proof of concept, we tested our methods on the Jahn-Teller induced \gls{MECI} of a smaller cluster, Ag$_3$.
As an initial geometry guess, we took one of the ground state minimum with bond lengths being $2.82$ \AA, $2.57$ \AA, and $2.57$ \AA.
The \gls{MECI} geometry is of D$_{3h}$ symmetry, and all methods, \gls{UBS}, \gls{ALM}, and \gls{SLM} converged to the same geometry with bond length $2.628$ \AA.
The convergence properties and the energies' evolution with respect to the number of iterations are presented in Tab.~\ref{tab:CASs} and in Fig.~\ref{fig:Ag3-energies}.
They show that all three methods converge quickly to the same \gls{MECI} with a similar degree of convergence, the \gls{ALM} method being the slowest to converge in this case.

%%%%%%%%%%%%%%%%%%%%%%%%%%%%%%%%%%%%%%%%%%%%%%%%%%%%%%%%%%%%%%%%%%%%%%
\section{Concluding remarks}
\label{sec:conclusion}

We present two new algorithms to optimize \gls{MECI} points.
These algorithms, denoted \gls{ALM} and \gls{SLM} in the text, employ the Lagrange multiplier methodology.
They converge to an exact \gls{CI} minimum and can be combined with a quasi-Newton optimization.
Hence, only energies and energy gradients are required to converge with an efficiency similar to the original well-known Lagrange multiplier method~\cite{Manaa:1993/jcp/5251} (referred as \gls{LM} in the main text).

We have shown the ability of presented methods (\gls{ALM} and \gls{SLM}) to converge to the \gls{MECI} on several molecular systems, including a large system of interest in bioluminescence, the furimamide, for which we obtained a first optimized \gls{MECI}, and a metallic cluster, Ag$_3$.

Our paper presents a basic implementation of the two methods, but could definitely be improved using the usual technics developed to improve the convergence of molecular geometry optimizations such as employing convergence accelerator,~\cite{Schlegel:2011/wcms/790} using internal redundant coordinates,~\cite{Schlegel:1984/tca/333,Pulay:1992/jcp/2856} or improving the initial guess of the Hessians.
Another point of possible improvement comes from our assumption that the electronic structure problem is solved, but it is known that exact forces can differ from their expression from the Hellmann-Feynman theorem due to the basis dependence to the nuclear coordinates, giving rise to the so-called Pulay terms.~\cite{Pulay:1969/mp/197}
Hence, one should test the importance of these terms and include them from the start in App.~\ref{app:ddO2}.
Finally, we should point out that the approximate construction of the \gls{DC} relies on the fitting of an underlying diabatic model, which can be greatly improved by employing machine learning techniques.~\cite{Chen:2023/mol/4222}

While the \gls{ALM} method is an application of the original and efficient \gls{LM} method, the approximate construction of the \gls{DC} can prevent from a quick convergence when the initial geometry is far from a \gls{CI}.
On the contrary, the \gls{SLM} method does not rely on an approximate calculation of the \gls{DC}, and seems to be a good alternative far from the \gls{CI}, but was found to be less efficient once the \gls{IS} is reached.
Thus, a natural future investigation is to explore combinations of the two methods depending on the energy difference.

%%%%%%%%%%%%%%%%%%%%%%%%%%%%%%%%%%%%%%%%%%%%%%%%%%%%%%%%%%%%%%%%%%%%%%
\section{Acknowledgments}

LJD thanks Benjamin Lasorne and Joachim Galiana for fruitful discussions and critical reading of the manuscript.
The author acknowledges financial support from French National Research Agency through the project ANR-21-CE29-0005.

%%%%%%%%%%%%%%%%%%%%%%%%%%%%%%%%%%%%%%%%%%%%%%%%%%%%%%%%%%%%%%%%%%%%%%
\appendix

%%%%%%%%%%%%%%%%%%%%%%%%%%%%%%%%%%%%%%%%%%%%%%%%%%%%%%%%%%%%%%%%%%%%%%
\section{Hessian of the energy difference squared}
\label{app:ddO2}

Assume that the electronic problem is solved, we can write the following relation
\bea
\op H_e \ket{\varphi_k} & = & E_k \ket{\varphi_k}.
\eea
Starting from  $\braOket{\varphi_k}{\mat\nabla\mat\nabla^t\Big[(\op H_e-E_k)}{\varphi_k}\Big]=0$, we can obtain the following relation for second derivative of the adiabatic energies:
\bea\label{eq:HessE}
\frac{\partial^2E_k}{\partial X_\alpha\partial X_\beta} & = & \braOket{\varphi_k}{\frac{\partial^2 \op H_e}{\partial X_\alpha\partial X_\beta}}{\varphi_k} \nonumber\\
&&+\sum_{l\ne k}\frac{\braOket{\varphi_k}{\frac{\partial \op H_e}{\partial X_\alpha}}{\varphi_l}\braOket{\varphi_l}{\frac{\partial \op H_e}{\partial X_\beta}}{\varphi_k}}{E_k-E_l} \nonumber\\
&&+\sum_{l\ne k}\frac{\braOket{\varphi_k}{\frac{\partial \op H_e}{\partial X_\beta}}{\varphi_l}\braOket{\varphi_l}{\frac{\partial \op H_e}{\partial X_\alpha}}{\varphi_k}}{E_k-E_l}. \nonumber\\
\eea
The Hessian of the squared adiabatic energy difference is
\bea\label{eq:HessDE2}
\frac{\partial^2[E_l-E_k]^2}{\partial X_\alpha\partial X_\beta} & = & 2[E_l-E_k]\frac{\partial^2[E_l-E_k]}{\partial X_\alpha\partial X_\beta} \nonumber\\
&&+2\frac{\partial[E_l-E_k]}{\partial X_\alpha}\frac{\partial[E_l-E_k]}{\partial X_\beta}.
\eea

Combining \eq{eq:HessDE2} and \eq{eq:HessE}, one arrives to the relation given in \eq{eq:graal_ahah}.
\begin{widetext}
\bea\label{eq:graal_ahah}
&&\frac{1}{2}\frac{\partial^2[E_l-E_k]^2}{\partial X_\alpha\partial X_\beta}-2\frac{\partial[E_l-E_k]}{\partial X_\alpha}\frac{\partial[E_l-E_k]}{\partial X_\beta} 
= 
\nonumber\\
&&
 [E_l-E_k] \Biggl\{
 \braOket{\varphi_l}{\frac{\partial^2 \op H_e}{\partial X_\alpha\partial X_\beta}}{\varphi_l} 
-\braOket{\varphi_k}{\frac{\partial^2 \op H_e}{\partial X_\alpha\partial X_\beta}}{\varphi_k} 
\nonumber\\
&&
+\sum_{m\ne l}\frac{\braOket{\varphi_l}{\frac{\partial \op H_e}{\partial X_\alpha}}{\varphi_m}\braOket{\varphi_m}{\frac{\partial \op H_e}{\partial X_\beta}}{\varphi_l}+\braOket{\varphi_l}{\frac{\partial \op H_e}{\partial X_\beta}}{\varphi_m}\braOket{\varphi_m}{\frac{\partial \op H_e}{\partial X_\alpha}}{\varphi_l}}{E_l-E_m} 
\nonumber\\
&&
-\sum_{m\ne k}\frac{\braOket{\varphi_k}{\frac{\partial \op H_e}{\partial X_\alpha}}{\varphi_m}\braOket{\varphi_m}{\frac{\partial \op H_e}{\partial X_\beta}}{\varphi_k}+\braOket{\varphi_k}{\frac{\partial \op H_e}{\partial X_\beta}}{\varphi_m}\braOket{\varphi_m}{\frac{\partial \op H_e}{\partial X_\alpha}}{\varphi_k}}{E_k-E_m} 
\Biggl\}
\nonumber
\eea
\end{widetext}
The equation~\ref{eq:graal_ahah} is exact.
When $E_k=E_l$, and if real electronic states are considered, \eq{eq:graal_ahah} reduces to \eq{eq:ddO2} such that \eq{eq:ddO2} is exact at the \gls{CI} geometry.
However, the second order terms containing $\mat\nabla\mat\nabla^t\op H_e$ as well as the terms involving other electronic states are negligible in the vicinity of a \gls{CI} because they are factored by $E_k-E_l\approx 0$.
Furthermore, even far from a \gls{CI}, if the ratio $[E_l-E_k]/[E_l-E_m]$ is small (other electronic states are far in energy from state $k$ and $l$) then the effect of extra electronic states is negligible as well.
Hence, \eq{eq:ddO2} can be used in the vicinity of \glspl{CI}.

%%%%%%%%%%%%%%%%%%%%%%%%%%%%%%%%%%%%%%%%%%%%%%%%%%%%%%%%%%%%%%%%%%%%%%
\section{Fit of the derivative coupling}
\label{app:fit}

To solve the equation $\mat Z(\mat p)=\mat 0$ where $\mat Z$ and $\mat p$ are defined in the main text in Eqs.~(\ref{eq:defZ}-\ref{eq:defp}), we first try the usual Newton step to update the set of parameters:
\bea\label{eq:Newtonp}
\mat p_{i+1} = \mat p_i - \mat J_i^{-1} \mat Z_i.
\eea
Here, the index $i$ indicates that the quantities are being calculated for the set of parameters $\mat p_i$, and $\mat J_i$ is the Jacobian matrix defined by
\bea
[\mat J_i]_{\alpha\beta} & = & \frac{\partial [\mat Z]_\alpha}{\partial p_\beta}\Big|_{\mat p_i}.
\eea
We evaluate the Jacobian numerically using finite central differences.
In \eq{eq:Newtonp}, the inverse of the Jacobian matrix is understood as a pseudo inversion using a singular value decomposition.

In some cases, $\norm{\mat Z_{i+1}}>\norm{\mat Z_{i}}$ such that the Newton step does not result in a minimization of the error.
If this occurs, we switch the algorithm to a trust-radius optimization, where the radius is progressively reduced until the step gives a minimization of the error.
This is achieved by minimizing the error with an added penalty on the step norm $\kappa\norm{ \mat p_{i+1}-\mat p_i }$ where $\kappa$ is a weight to be chosen.
We further approximate the Hessian of the error in the parameter space as $\mat J^t\mat J$, i.e. we assume that the term $\mat Z^t\partial^2\mat Z/(\partial p_{\alpha}\partial p_{\beta})$ becomes negligible at convergence, when the $\mat Z$ is presumably small enough.
The step is then given by
\bea\label{eq:LevMarqp}
\mat p_{i+1} = \mat p_i - ( \mat J_i^t\mat J_i + \kappa^2 \mat 1_{2\mathcal{D}+1} )^{-1} \mat J_i^t\mat Z_i.
\eea
This algorithm is in fact the Levenberg-Marquardt~\cite{Levenberg:1944/qam/164,Marquardt:1963/jsiam/431} revisited.
Equation~\ref{eq:LevMarqp} is repeatedly used with increasing values of $\kappa$ until the error satisfies $\norm{\mat Z_{i+1}}<\norm{\mat Z_{i}}$.
When trust radius is switched on, the initial value for the weight is set to $\kappa=10^{-3}$, and is increased by a factor of $10$ at each iteration ($10\times\kappa\to\kappa$).

The initial guess is chosen as $c=\Omega_n$, $\mat v=\mat\nabla\Omega_n$, and $\mat w=\mat\nabla\Omega_{n-1}$.

\bibliography{noNACoptCI}
\end{document}